\documentclass[11pt]{article}
\pdfoutput=1
\usepackage{amsmath}
\usepackage{amssymb}
\usepackage{graphicx,bbm,mathrsfs}
\usepackage{nicefrac}
\usepackage{slashed}
\usepackage{bbm}
\usepackage{geometry}
\usepackage{empheq}
\usepackage{stackrel}
\usepackage{jheppub}
\usepackage{setspace}
\usepackage{relsize}
\usepackage{empheq}
\usepackage{pifont}
\usepackage{mathtools}
\usepackage{enumitem}
\usepackage{dcolumn}   
\usepackage{bm}        
\usepackage{graphicx,mathrsfs}
\usepackage{nicefrac}
\usepackage{multirow}
\usepackage{color}
\usepackage{mathtools}
\usepackage{makecell}
\usepackage{environ} 
\usepackage{lipsum} 
\usepackage{wasysym}
\usepackage{hhline,colortbl}  
\usepackage{tikz}
\usepackage{graphicx}
\usepackage{tensor}
\usepackage{dsfont}
\usepackage{simpler-wick}
\usepackage{accents}
\usepackage{soul}

\usepackage{booktabs}
\usepackage{mdframed}
\usepackage{xcolor}
\definecolor{EqFrame}{RGB}{235,245 ,250 }

 \NewEnviron{Smaller11}{
           \scalebox{1.1}{$\BODY$} 
 } 
 \NewEnviron{Smaller08}{
           \scalebox{0.8}{$\BODY$} 
 }

\newcommand{\be}{\begin{equation}}
\newcommand{\ee}{\end{equation}}
\newcommand{\bea}{\begin{eqnarray}}
\newcommand{\eea}{\end{eqnarray}}

\renewcommand\labelenumi{(\roman{enumi})}
\renewcommand\theenumi\labelenumi

\renewcommand{\Re}{\operatorname{Re}}
\renewcommand{\Im}{\operatorname{Im}}

\usepackage{titlesec}

\titleformat*{\section}{\Large\bfseries}
\titleformat*{\subsection}{\large\bfseries}
\titleformat*{\subsubsection}{\large\bfseries}
\titleformat*{\paragraph}{\large\bfseries}
\titleformat*{\subparagraph}{\large\bfseries}

\makeatletter
\newcommand*{\prodsym}{%
  \DOTSB
  \mathop{
    \mathchoice
      {\rlap{\kern.3em\rotatebox[origin=c]{-90}{}}{\prod}}
      {\vcenter{\rlap{\kern.2em\rotatebox[origin=c]{-90}{}}}{\prod}}
      {\sum}{\sum}
  }\slimits@
}
\makeatother

\makeatletter
\DeclareFontFamily{OMX}{MnSymbolE}{}
\DeclareSymbolFont{MnLargeSymbols}{OMX}{MnSymbolE}{m}{n}
\SetSymbolFont{MnLargeSymbols}{bold}{OMX}{MnSymbolE}{b}{n}
\DeclareFontShape{OMX}{MnSymbolE}{m}{n}{
    <-6>  MnSymbolE5
   <6-7>  MnSymbolE6
   <7-8>  MnSymbolE7
   <8-9>  MnSymbolE8
   <9-10> MnSymbolE9
  <10-12> MnSymbolE10
  <12->   MnSymbolE12
}{}
\DeclareFontShape{OMX}{MnSymbolE}{b}{n}{
    <-6>  MnSymbolE-Bold5
   <6-7>  MnSymbolE-Bold6
   <7-8>  MnSymbolE-Bold7
   <8-9>  MnSymbolE-Bold8
   <9-10> MnSymbolE-Bold9
  <10-12> MnSymbolE-Bold10
  <12->   MnSymbolE-Bold12
}{}

\let\llangle\@undefined
\let\rrangle\@undefined
\DeclareMathDelimiter{\llangle}{\mathopen}%
                     {MnLargeSymbols}{'164}{MnLargeSymbols}{'164}
\DeclareMathDelimiter{\rrangle}{\mathclose}%
                     {MnLargeSymbols}{'171}{MnLargeSymbols}{'171}
\makeatother

\begin{document}

\vspace*{4mm}

\thispagestyle{empty}

\begin{center}

\begin{minipage}{20cm}
\begin{center}
\hspace{-6cm }
\huge
\sc
Massive Graviton Dark Matter  \\ \hspace{-6cm}from a Gapped Continuum 
\end{center}
\end{minipage}
\\[30mm]

\renewcommand{\thefootnote}{\fnsymbol{footnote}}

{\large\bf
Eugenio~Meg\'{\i}as$^{\, a}$~\footnote{emegias@ugr.es}\,,
Miguel~Prieto$^{\, b}$~\footnote{miguel.prfdez@gmail.com}\,, 
Mariano~Quir\'os$^{\, c}$~\footnote{quiros@ifae.es}\,
}\\[12mm]
\end{center} 
\noindent

${}^a\!$ 
\textit{Departamento de F\'{\i}sica At\'omica, Molecular y Nuclear and Instituto Carlos I de F\'{\i}sica Te\'orica y Computacional, Universidad de Granada, Avenida de Fuente Nueva 
s/n, 18071 Granada, Spain}

${}^b\!$ 
\textit{Departament de F\'{\i}sica
Qu\`{a}ntica i Astrof\'{\i}sica, Universitat de Barcelona, Mart\'{\i} i Franqu\`{e}s 1, 08028 Barcelona, Spain}

${}^c\!$  
\textit{Institut de F\'{\i}sica d'Altes Energies (IFAE) and
The Barcelona Institute of  Science and Technology (BIST), Campus UAB, 08193 Bellaterra, Barcelona, Spain}

\vspace*{8mm}
 
\begin{center}
{  \bf  Abstract }
\end{center}
\begin{minipage}{16cm}
\setstretch{0.95}
We consider the possibility of dark matter in a warped extra-dimensional theory in the presence of a linear dilaton background, with a gapped continuum spectrum, in a brane-world cosmological scenario. Firstly, triggered by self-energy radiative corrections, we study the existence of an isolated resonance of massive gravitons, and its realization as a long-lived feebly interacting dark matter candidate, produced by the freeze-in mechanism. This massive graviton is proved to satisfy all theoretical and experimental constraints, in the sub-MeV mass range. We further consider the close relationship between the existence of this component of dark matter and the presence of an inflaton localized on the brane, with a mass around $10^{11}$~GeV and a sub-TeV reheating temperature, in a brane inflationary scenario that allows one to reproduce the most recent cosmological observables. Secondly, the gapped continuum of gravitons, a particular five-dimensional realization of the physics of unparticles, is identified as a holographic fluid which can play the role of holographic dark matter. The production of the holographic fluid goes by an ultraviolet freeze-in mechanism, with an abundance mainly depending on the reheating temperature. Depending on the values of the mass gap and the reheating temperature, one or both components of dark matter can be present.
\end{minipage}

\newpage
\setcounter{tocdepth}{2}

\tableofcontents  

\vspace{1cm}
\hrule
\vspace{1cm}

\newpage

\section{Introduction}

A warped extra dimension in five-dimensional (5D) space was introduced
by Randall and Sundrum, Ref.~\cite{Randall:1999ee}, to solve the
hierarchy problem in particle physics. The 5D space was endowed with
an anti-de Sitter (AdS) bulk background and two three-branes, one in
the ultraviolet (UV) with a cutoff equal to the four-dimensional (4D)
reduced Planck scale $M_4=2.43\times 10^{18}$~GeV, and the other in
the infrared (IR) with a cutoff at the TeV scale. For the case of AdS
background, there is the additional bonus that the bulk 5D theory is
dual to a 4D holographic theory, a conformal field theory (CFT) in the
nonperturbative regime.

Moreover, it was shown in Ref.~\cite{Randall:1999vf} that, in an AdS
construction with an infinite extra dimension and the matter in a
brane, the extra dimension needs not be small, as 5D gravity is acting
in the three-brane as 4D general relativity. These constructions were
motivated by string theories, for which the open string sector
describing matter fields was living on the lower dimensional brane,
while the closed string sector, related to the gravitational modes,
was propagating in the bulk of the whole space. Due to the underlying
conformal symmetry of the theory, the bulk graviton propagator is an
ungapped continuum from which the zero mode is identified with the
physical graviton, i.e.~a particular 5D modeling of
unparticles~\cite{Georgi:2007ek}.  These theories are dubbed
world-brane models as our observable universe is confined to a
three-brane while gravity, or other fields, can propagate into the
extra dimension. It was also
proved~\cite{Shiromizu:1999wj,Binetruy:1999hy} that, from the 5D
Einstein's equations with a constant potential (cosmological constant)
as a source in the bulk, the cosmology in the brane is different from
the conventional 4D cosmology, as the Hubble parameter which defines
the cosmological evolution of the matter fields, has two extra pieces
\be
3 H^2 M_4^2=\rho_b+\frac{\rho_b^2}{2\lambda}+\rho_{\rm fluid} \,,\quad \lambda=6 M_5^6/M_4^2 \,,
\label{eq:H2brane}
\ee
 where $M_5$ is the 5D Planck scale, and $\rho_b$ is the energy density of matter in the brane. The second term in (\ref{eq:H2brane}) is only relevant for cosmology at very high energy ($\rho_b\gg 2\lambda$), while the third one describes a dark radiation fluid as it scales with the scale factor $a$ as $\rho_{\rm fluid}^{\rm AdS}\sim 1/a^4$~\cite{Binetruy:1999hy,Hebecker:2001nv,Langlois:2002ke,Langlois:2003zb}.
 
 A different theory where the bulk graviton propagator is a gapped continuum appeared in a different class of 5D backgrounds, the so-called linear dilaton (LD)~\cite{Antoniadis:2011qw,Cox:2012ee} metric, which is known to be dual to little string theories~\cite{Aharony:1999ks,Antoniadis:2001sw}. The metric is induced by an exponential potential in the bulk with a critical exponent and was shown in Ref.~\cite{Cabrer:2009we} to lead, on top of the massless physical graviton pole, to a gapped continuum graviton propagator. It was also proved, by solving the 5D Einstein equations with the critical bulk potential leading to the LD background, that the cosmology in the brane is different from the dark radiation characteristic of an AdS background. In particular it provides, for the last term in Eq.~(\ref{eq:H2brane}), a matter behaved fluid, $\rho_{\rm fluid}^{\rm LD}\sim 1/a^3$~\cite{Fichet:2022xol,Fichet:2023dju,Fichet:2023xbu}, which could play the role of holographic dark matter (DM), as was shown in Ref.~\cite{Fichet:2026nct}.
 
 In this paper we will explore another possible DM candidate in 5D theories in the presence of the LD background. In fact, as the continuous graviton distribution is coupled to the energy-momentum tensor of the Standard Model fields, the renormalization of the graviton self-energy can give rise to an isolated resonance, a long-lived massive graviton which we have shown to have the capabilities to make the DM of the universe. It is an alternative, depending on the values of the parameters of the theory, to the holographic fluid, or even it is possible that both make part of the DM components. The appearance of an isolated resonance from a continuous distribution is a phenomenon which was known and studied long ago in the case of unparticles~\cite{Delgado:2008gj,Megias:2023kpk}. Moreover we have shown that the presence of an inflaton, with a mass $m\simeq M_5$, localized in the brane plays an essential role for the existence of an isolated graviton resonance with a mass near the mass gap of the continuous graviton propagator, while during inflation the second term in (\ref{eq:H2brane}) controls the expansion of the universe, characteristic of high-energy brane cosmology. We also have proposed a simple brane inflationary model reproducing the most recent cosmological observables.

 The manuscript is organized as follows. In section~\ref{sec:model} we introduce the 5D warped dimensional model characterized by a dilaton-gravity background, and obtain the graviton propagator. We study in section~\ref{sec:self_energy} the one-loop corrections to the massive graviton self-energy, which follows from the coupling of the graviton to the different portals: scalars, fermions and gauge bosons. The properties of the isolated resonance in the graviton propagator
 are studied in section~\ref{sec:longlived_DM} and the possibility that it can be a dark matter candidate in the context of a long-lived feebly interacting massive particle (FIMP) scenario. Section~\ref{sec:unparticles} is devoted to study the relation between the graviton propagator of the model, and the physics of unparticles. We review the fluid production by a freeze-in mechanism due to the leakage of gravitons from the brane into the bulk in section~\ref{sec:fluid}, and study, in section~\ref{sec:brane_inflation}, the consequences on the model of a class of inflationary scenarios controlled by a quadraticlike potential.   We present in section~\ref{sec:conclusion} our conclusions and an outlook for future directions. Finally, we provide in appendix~\ref{sec:appendix} some technical details on the one-loop corrections to the self-energy of the graviton.

\section{The model}
\label{sec:model}

Our setup is a dilaton gravity 5D model with a 4D brane characterized by the action in the Einstein frame given by
\begin{align}
\mathcal S&=M_5^3\int d^5x \sqrt{g}\left(\frac{1}{2}R_5-\frac{3}{2}(\partial_M\bar\phi)^2 -3\bar V(\bar\phi)\right)\nonumber\\
&-\int_{b}d^4x\sqrt{\bar g}\left(V_b(\phi)+\Lambda_b-M_5^3 K-\mathcal L_{\rm matter}\right)  \,,
\end{align} 
where $R_5$ is the 5D scalar curvature, $M_5$ the fundamental 5D Planck scale, $\phi\equiv \sqrt{3M_5^3}\,\bar \phi$ the dilaton field, $K$ the extrinsic curvature, and $V\equiv 3 M_5^3 \,\bar V$ the bulk potential. The brane, with induced metric $\bar g_{\mu\nu}$, supports a tension $\Lambda_b$ and a localized potential $V_b(\phi)$ which does not need to be specified as it fixes, in the stiff limit, the vacuum expectation value (VEV) of $\langle \phi\rangle= v_b$ and such that $V_b(v_b)=0$. 

We will consider in the following the bulk potential
\begin{equation}
\bar V(\bar \phi) = -\frac{3}{2} k^2 e^{2 \bar\phi} \,,  \label{eq:Vb}
\end{equation}
where $k$ is a constant with dimension of mass.

\subsection{The holographic fluid}
\label{subsec:fluid}

A black hole solution of the 5D Einstein field equations with the potential of Eq.~(\ref{eq:Vb}) is given in the brane cosmology coordinates by
\begin{eqnarray}
  ds^2 &=& g_{MN} dx^M dx^N =  \left( \frac{r}{L} \right)^2 \left( f(r) d\tau^2 - d \mathbf{x}^2  \right)   -  \frac{1}{(\eta r_b)^2 } \frac{dr^2}{f(r)} \,,  \label{eq:ds2BH}\\
  \bar\phi(r) &=&  \bar v_b - \log\left( \frac{r}{r_b} \right) \,, \qquad f(r) = 1 - \frac{r_h^3}{r^3} \,, \qquad \eta = k \, e^{\bar v_b} \,.  \label{eq:f}
\end{eqnarray}
The brane location is fixed at the position $r = r_b$, and $\bar v_b$ is the VEV of the scalar field at the brane. The metric~(\ref{eq:ds2BH}) is appropriate to describe cosmology as seen from the brane standpoint. The induced metric on the brane is
  \begin{equation}
  d\bar s^2 = \bar g_{\mu\nu} dx^\mu dx^\nu \equiv dt^2 - \frac{r_b^2}{L^2} d\mathbf x^2 \,, \label{eq:ds2FRW}
  \end{equation}
  where we have introduced the brane cosmic time $dt = \frac{r_b}{L} \sqrt{f(r_b)} d\tau$. If the brane moves along $r$ in the 5D background, the observer located at the brane perceives expansion of the 4D universe with Hubble parameter $H = \dot r_b / r_b$, where $\dot r_b \equiv \partial_t r_b$. Our convention is to choose $r_b$ equals $L$ at present times, such that $r_b = a(t) L$ where $a(t) \equiv a(r_b(t))$ is the standard scale factor. Then, Eq.~(\ref{eq:ds2FRW}) corresponds to the standard Friedmann-Lema\^{\i}tre-Robertson-Walker metric.

  By using Eq.~(\ref{eq:ds2BH}), the  physical black hole temperature measured by an observer on the brane is given by
\begin{equation}
    T_b = \frac{T_h}{a(r_b)} = \frac{3 \eta}{4\pi} \,,
\end{equation}
where $T_h  = \frac{3 \eta}{4\pi} \frac{r_b}{L}$ is the Hawking temperature of the black hole, i.e.~$T_b$ is related to the Hawking temperature by cosmological redshift. The entropy of the horizon is given by
\begin{equation}
\mathcal S_h =  \frac{\mathcal A}{2 G_5} \,,\qquad G_5=\frac{1}{8\pi M_5^2} \,,
\label{eq:sh0}
\end{equation}
where $G_5$ is the 5D Newton constant, and the area of the event horizon is computed~as
\begin{equation}
\mathcal A = \int d^3x \sqrt{|\bar g|} = V_3 \left( \frac{r_h}{L} \right)^3 \,.
\end{equation}
The entropy density is then given by
\begin{equation}
s_h(r_h,r_b) \equiv \frac{\mathcal S_h}{V_b} = 4 \pi M_5^3 \left( \frac{r_h}{r_b} \right)^3 \,,  \label{eq:sh}
\end{equation}
where $V_b = V_3 a^3(r_b)$ is the spatial volume of the brane, with $V_3 = \int d^3x$ the comoving volume.

The physics of a brane-localized observer is governed by the effective Einstein equation
\begin{equation}
{}^{(4)} G_{\mu\nu} = \frac{1}{M_4^2} \left(  T_{\mu\nu}^b + T_{\mu\nu}^{\textrm{eff}}\right) + O\left( \frac{T_b^2}{M_5^6} \right) \,,
\end{equation}
where $T_ {\mu\nu}^b$ is the stress tensor  of the brane-localized matter, and $T_{\mu\nu}^{\textrm{eff}}$ is the bulk energy-momentum tensor projected on the brane~\cite{Shiromizu:1999wj,Fichet:2022ixi,Fichet:2022xol}. We are working in the low-energy limit where $|T^b_{\mu\nu}|  \ll M_5^6 / M_4^2$. Following~\cite{Fichet:2023xbu}, the projection procedure for the $T_{\mu\nu}^{\textrm{eff}}$ tensor leads to the structure of a four-dimensional perfect fluid at rest given by
\begin{equation}
T_{\nu}^{\textrm{eff} \,, \mu} = g^{\mu\lambda} T_{\lambda\nu}^{\textrm{eff}} = \textrm{diag}(-\rho_h, P_h, P_h, P_h) \,.
\end{equation}
We refer to this as the holographic fluid. 

The effective energy density and pressure split as
\begin{equation}
\rho_{h} = \rho_{\textrm{fluid}} + \rho_{\textrm{vacuum}} \,, \qquad P_{h} = P_{\textrm{fluid}} + P_{\textrm{vacuum}} \,,
\end{equation}
and the vacuum contribution is given by
\begin{equation}
  \rho_{\textrm{vacuum}} = - P_{\textrm{vacuum}} = \Lambda_4 M_4^2  \,,
\end{equation}
where the 4D cosmological constant $\Lambda_4$ and the Planck mass $M_4$ are related by
\begin{equation}
M_5^3 = \eta M_4^2 \sqrt{ 1 + \frac{\Lambda_4}{3\eta^2} } \,,
\end{equation}
and the fluid contributions turn out to be
\begin{equation}
\rho_{\textrm{fluid}} = 3 \eta^2 M_4^2 \left( \frac{r_h}{r_b} \right)^3 \,, \qquad P_{\textrm{fluid}} = 0 \,. \label{eq:rhoh}
\end{equation}
The brane tension can be ultimately tuned to set $\Lambda_4$ to (almost) zero, so that we will consider that the 5D Planck mass turns out to be related to the 4D one by $M_5^3=\eta M_4^2$. Then we obtain the first Friedmann equation on the brane, in the low-energy limit, as 
\begin{equation}
3M_4^2 H^2 = \rho_b  + O\left( \frac{\rho_b^2}{\eta^2 M_4^2} \right) + \rho_{\textrm{fluid}}\,.
\end{equation}

An alternative way to obtain this result is by considering the fundamental law of thermodynamics
\begin{equation}
dE_{\textrm{fluid}} = T_b \, d\mathcal S_{\textrm{fluid}} - P_{\textrm{fluid}} \, dV_b \,.
\end{equation}
By using $\mathcal S_{\textrm{fluid}} = \mathcal S_h$ given by Eq.~(\ref{eq:sh}), and the condition of pressureless fluid, $P_{\textrm{fluid}} = 0$, one gets the energy density of the holographic fluid $\rho_{\textrm{fluid}} = E_{\textrm{fluid}} / V_b$ given by Eq.~(\ref{eq:rhoh}), which turns out to be equal to $T_b \, s_{\textrm{fluid}}(r_h) $.

The continuity equations for the brane and fluid energy densities, $\rho_b$ and $\rho_{\textrm{fluid}}$, are given~by
\begin{eqnarray}
  \dot \rho_b &=& -4 H \rho_b +  \Delta_b \,, \\
    \dot \rho_{\textrm{fluid}} &=& - 3 H \rho_{\textrm{fluid}}  + \Delta_{\textrm{fluid}} \,, \qquad   \Delta_b + \Delta_{\textrm{fluid}} = 0 \,,
\end{eqnarray}
where the emissivity~\cite{Fichet:2026nct}
\begin{equation}
\Delta_{\textrm{fluid}} = c \frac{T^8}{\eta M_4^2} \,, \qquad \textrm{with} \qquad c \simeq 3.919 \,,
\end{equation}
is triggered by the leakage of gravitons from the brane into the bulk.

These equations can be easily solved by considering the relation $d/dt  = - H T d/dT$. When neglecting the leakage of gravitons $(c \to 0)$, one gets the solution
\begin{equation}
\rho_{\textrm{fluid}}(T) \propto T^3 \qquad \Longrightarrow \qquad \frac{r_h}{r_b} \propto T \,.
\end{equation}
On the other hand, when considering the leakage of gravitons $(c \ne 0)$, one gets the following solution
\begin{equation}
\rho_{\textrm{fluid}}(T) = - \left( \frac{10}{g_{\textrm{eff}}}  \right)^{1/2} \frac{c}{\pi \eta M_4} T^6  + C_h T^3 =   \left( \frac{10}{g_{\textrm{eff}}}  \right)^{1/2} \frac{c}{\pi \eta M_4} T^3 \left( T_R^3  - T^3 \right) \,,  \label{eq:rhoh2}
\end{equation}
where $C_h$ is an integration constant that has been fixed by the condition that the energy density of the holographic fluid is vanishing at the reheating temperature, i.e.~$\rho_{\textrm{fluid}}(T_R) = 0$. For $ T \ll T_R$ the solution is
\begin{equation}
\rho_{\textrm{fluid}}(T) \simeq \left( \frac{10}{g_{\textrm{eff}}}  \right)^{1/2} \frac{c}{\pi \eta M_4} T_R^3 T^3   \,.
\end{equation}

Notice that the continuity equation for $\rho_{\textrm{fluid}}$ can be expressed as an equation for $r_h(T)$ when using the expression for the pressureless fluid, Eq.~(\ref{eq:rhoh}), leading to
\begin{equation}
r_h^\prime(T) + \frac{c \sqrt{10}}{3 \pi \sqrt{g_{\textrm{eff}}}} \left(\frac{r_b(T)}{\eta M_4}\right)^3  \frac{T^5}{r_h^2(T)} = 0  \,.
\end{equation}
In getting this equation we have used that $r_b(T) \propto 1 / T$. The solution of this equation is
\begin{equation}
\left(\frac{r_h}{r_b}\right)^3 =  \frac{c \sqrt{10}}{3\pi \sqrt{g_{\textrm{eff}}}} \left( \frac{1}{\eta M_4} \right)^3 T^3 (T_R^3 - T^3) \,,
\end{equation}
which satisfies the boundary condition $r_h(T_R)=0$ so that the final value is obtained by UV freeze-in. For $T\ll T_R$, the solution is
\be
\left(\frac{r_h}{r_b}\right)^3 =  \frac{c \sqrt{10}}{3\pi \sqrt{g_{\textrm{eff}}}} \left( \frac{1}{\eta M_4} \right)^3 T_R^3 \,T^3\,. 
\ee
It is straightforward to check that this solution leads to Eq.~(\ref{eq:rhoh2}) when plugging it into the expression of $\rho_{\textrm{fluid}}(r_h)$ given by Eq.~(\ref{eq:rhoh}), i.e.
\begin{equation}
\rho_{\textrm{fluid}}(T) = \rho_{\textrm{fluid}}(r_h(T),r_b(T)) \,,
\end{equation}
as expected.

\subsection{The graviton propagator}
\label{subsec:graviton_propagator}

We will now study the graviton propagator in the absence of black hole. The use of this propagator will be justified in section~\ref{sec:fluid} based on the fact that the horizon becomes very far away from the brane for the typical values of the parameters of the model.

The brane at $r = r_b$ partitions the space-time $\mathcal M$ into two regions
\begin{equation}
\mathcal M^- \equiv \mathcal M |_{(0,r_b]} \,, \qquad \mathcal M^+  \equiv \mathcal M |_{[r_b, \infty)} \,.
\end{equation}
As we will see below, one of the differences between these subspaces is the presence or absence of graviton zero mode.

The system can also be described in conformal coordinates $z$, with the solution~\footnote{We denote the 4D comoving coordinates as $x^\mu$ and the physical ones as $\tilde x^{\mu}$. Note that the relation between comoving and physical coordinates is $
\tilde x^\mu=a x^\mu,\; \tilde x_\mu=\frac{1}{a}x_\mu
$,
where we have used that comoving (physical) coordinates change with the metric $g_{\mu\nu}$  ($\eta_{\mu\nu}$). Therefore, while comoving coordinates with contravariant indices are conserved, comoving coordinates with covariant indices are not, as $x_\mu=a^2 \eta_{\mu\nu}x^\nu$. In addition, the condition $\tilde x^2=x^2$ is satisfied. Finally note that the fifth coordinate is also physical, $z=\tilde x^5$.

Similarly the conserved comoving momenta are $p_\mu$, with covariant indices. Comoving momenta with contravariant indices are not conserved as $p^\mu=\frac{1}{a^2}\eta^{\mu\nu}p_\nu$. Comoving momenta transform with the metric $g_{\mu\nu}$. Physical momenta are then $\tilde p_\mu=\frac{1}{a}p_\mu$ and $\tilde p^\mu=a p^\mu$ and transform with the metric $\eta_{\mu\nu}$. For the quadratic invariant they satisfy the condition $\tilde p^2=p^2$. Notice finally that comoving momenta are tied to the expanding coordinate grid, while physical momenta are what a local observer would actually measure with a detector.
}
\begin{eqnarray}
  ds^2 &=& e^{-2A(z)} \left( \eta_{\mu\nu} d\tilde x^\mu d\tilde x^\nu  - dz^2 \right)  \,,  \label{eq:ds2z}\\
 A(z) &=& \eta (z - z_b)  \,, \qquad  \bar\phi(z) =  \bar v_b + \eta (z  - z_b)  \,, \qquad \eta = k \, e^{\bar v_b} \,,  \label{eq:ds2z2}
\end{eqnarray}
where $\tilde x^\mu = a(z_b) x^\mu$ are the 4D physical coordinates, with $a(z_b) = e^{-\eta z_b}$ the scale factor. The relation with the brane cosmology coordinates is $\eta z=-\log(r/L)$. The brane position at $r=r_b$ then corresponds to $z_b= - \frac{1}{\eta} \log(r_b / L)$, while the singularity at $r_s=0$ corresponds to $z_s=\infty$. In this way the space $\mathcal M^-$  corresponds to the interval $z\in[z_b,\infty)$, while the space $\mathcal M^+$ to the interval $z\in(-\infty,z_b]$.~\footnote{In the presence of a black hole, the horizon is in the $\mathcal M^-$ space, at $z_h=- \frac{1}{\eta} \log(r_h/L)$, and the distance between the brane and the horizon behaves in terms of the temperature $T$ as $z_h(T) - z_b(T) \simeq - \frac{1}{\eta}\log(T_R T/(\eta M_4))$. Then, the horizon gets further and further away from  the brane as $T\to 0$.}

The graviton is a transverse traceless fluctuation of the metric of the form
\begin{equation}
ds^2 = e^{-2A(z)} \left[ \left(\eta_{\mu\nu} + \frac{2}{M_5^{3/2}} h_{\mu\nu}(\tilde x,z) \right) d\tilde x^\mu d\tilde x^\nu - dz^2  \right] \,,
\end{equation}
where we are considering conformal coordinates. If one works with the rescaled field defined by $\tilde h_{\mu\nu}(\tilde x,z) = e^{-3A(z)/2} h_{\mu\nu}(\tilde x,z)$, the corresponding 5D graviton propagator in the subspace $\mathcal M^\pm$ is given by
\be
G^\pm_{\mu\nu;\alpha\beta}(z,z^\prime;p^2) = G^\pm_h(z,z^\prime;p^2) P_{\mu\nu;\alpha\beta}(p^2) \,,
\ee
where
\begin{equation}
P_{\mu\nu;\alpha\beta}(p^2) = \frac{1}{2}\left(\Theta_{\mu\alpha}\Theta_{\nu\beta}+\Theta_{\mu\beta}\Theta_{\nu\alpha} \right)-\frac{1}{3}\Theta_{\mu\nu}\Theta_{\alpha\beta},\quad \Theta_{\mu\nu}=\eta_{\mu\nu}-\frac{p_\mu p_\nu}{p^2}  \,,
\end{equation}
while the coupling of the graviton $h_{\mu\nu}$ with the SM fields is driven by the conserved energy-momentum tensor $T^{\mu\nu}_{\rm SM}$, i.e.
\be
\mathcal L_{\textrm{h-matter}}=\frac{1}{M_5^{3/2}}\,h_{\mu\nu}(z_b,\tilde x)T^{\mu\nu}_{\rm SM}(\tilde x)\,. \label{eq:couplinghT}
\ee

After solving the equations of motion of the graviton with Neumann boundary condition on the brane $\partial_z G(z,z^\prime)|_{z = z_b} = 0$, and regularity of the solution in the limit $z \to \pm \infty$, one finds that the scalar part of the 5D brane-to-brane graviton propagator is given by~\cite{Megias:2021mgj,Fichet:2026nct}
\be
G_{h}^\pm(z_b,z_b;s)= - \frac{ 1 }{m_g} \frac{1}{\Delta \pm 1} \,,
\ee
where
\begin{equation}
\Delta = \sqrt{1 - \frac{s}{m_g^2}} \,, \qquad \left(s \equiv p^2,\ m_g=\frac{3}{2}\eta\right) \,.
\end{equation}
There is a continuum of KK graviton modes for $m > m_g$, where $m_g $ is the mass gap. For $\mathcal M^-$ there is also a massless mode corresponding to the physical graviton in 4D (this massless mode is absent in the $\mathcal M^+$ subspace). This can be seen by computing the limit of the propagator for $s\to 0$, for which one gets
\be
G_{h}^-(z_b,z_b;s) \stackrel[ s \to 0 ]{}{\simeq} \frac{2 m_g }{s} \,,
\label{eq:graviton}
\ee
while for $G_{h}^+$ there is a constant behavior. Remarkably, the graviton propagator in the $\mathcal M^-$ region with the zero mode subtracted coincides with the propagator in the $\mathcal M^+$ region,~i.e.
\begin{equation}
G_h(z_b,z_b;s) \equiv G_h^-(z_b,z_b;s) - \frac{2m_g}{s} = G_h^+(z_b,z_b;s) = -\frac{1}{m_g} \frac{1}{\Delta + 1} \,. \label{eq:Gh}
\end{equation}
In the following we will consider the propagator given by Eq.~(\ref{eq:Gh}) (which does not contain the massless graviton propagator), as the physical graviton mass is protected by diffeomorphism (general coordinate) invariance ($h_{\mu\nu}\to h_{\mu\nu}+\partial_\mu\xi_\nu+\partial_\nu \xi_\mu$), in much the same way as how gauge invariance protects the photon mass in QED.

Finally notice that we have used physical coordinates in the propagator calculation, so the momenta which appear are physical ones, i.e.~the momenta that would be measured by a local experiment. Still for computing physical processes, as cross sections or interaction rates $\Gamma_{\rm int}=n\langle \sigma v\rangle$, which are much larger than the Hubble $\Gamma_{\rm int}\gg H$, we can consider the spacetime is Minkowski, as the characteristic timescale is $\tau_{\rm int}\simeq 1/\Gamma_{\rm int}\ll \tau_H\simeq 1/H$, and the processes take place essentially in flat spacetime. Of course, the expansion of the universe is properly taken into account in the Boltzmann equations when computing cosmological quantities as number, or energy, densities.

\section{Self-energy from the different portals}
\label{sec:self_energy}

\begin{figure}[htb]
 \begin{center}
  \includegraphics[width=0.5\textwidth]{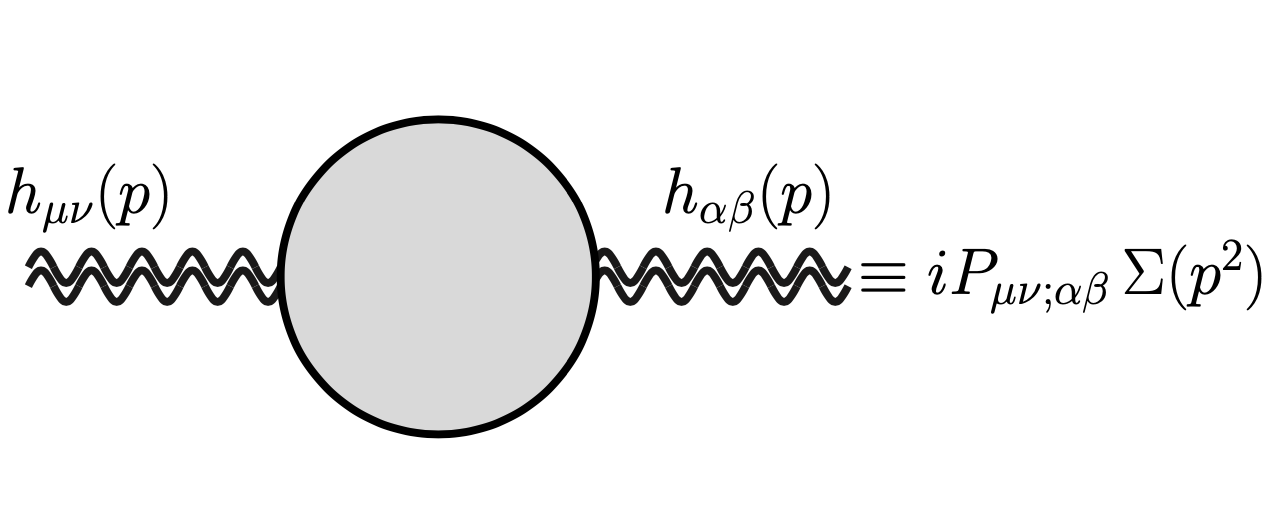} 
  \end{center}
\vspace{-0.7cm} \caption{\it Structure of the self-energy corrections}
     \label{fig:freeE}
 \end{figure}

 In this section we will compute the one-loop corrections to the massive graviton self-energy, from the different portals $\varphi\varphi$ coupled to the graviton by Eq.~(\ref{eq:couplinghT}). They have to respect the symmetries of the tree-level propagators, as indicated in Fig.~\ref{fig:freeE}. Technical details are postponed to App.~\ref{sec:appendix}.
 We will compute in all cases the contribution from sunset diagrams using the linear interaction $h\varphi\varphi$. In the case of scalars $\phi$ we will find it useful, as we will see, to also consider the contribution from the quadratic interaction $hh\phi\phi$ in seagull diagrams.

 \subsection{Scalars}
 As already stated, for the case of scalars we will consider the contribution from both, the sunset and seagull, diagrams, although the
  latter will only renormalize the location of the pole mass and not add anything in the complex $s$-plane, i.e.~to the pole width.

\subsubsection{Sunset diagram}

For a real scalar $\phi$ with mass $m$ the energy-momentum tensor is
\be
T_{\mu\nu}=\partial_\mu\phi\partial_\nu\phi-\eta_{\mu\nu}\left(\frac{1}{2}\eta^{\alpha\beta}\partial_\alpha\phi\partial_\beta\phi -\frac{1}{2}m^2\phi^2 \right)  \,.
\label{eq:tensor-scalar}
\ee
The Feynman rule for interaction of two scalars (with incoming momenta $p_1$ and $p_2$) with the graviton $h_{\mu\nu}$ [with incoming momentum $p=-(p_1+p_2)$]  is
\be
i\Gamma^{\mu\nu}(p_1,p_2)=i\kappa \left(p_1^\mu p_2^\nu+p_1^\nu p_2^\mu -\eta^{\mu\nu}\left(p_1\cdot p_2 + m^2 \right) \right) \,,\qquad \kappa\equiv M_5^{-3/2} \,.
\label{eq:coupling-scalar}
\ee

The structure of the dressed propagator $\bar G_{\mu\nu;\alpha\beta}(p)$ is given by
\be
\bar G_{\mu\nu;\alpha\beta}(p) = \frac{i P_{\mu\nu;\alpha\beta}}{G_h^{-1}}+\frac{i P_{\mu\nu;\alpha'\beta'}}{G_h^{-1}}(-i\Sigma^{\alpha'\beta';\mu'\nu'})\frac{i P_{\mu'\nu';\alpha\beta}}{G_h^{-1}} + \cdots \,,
\label{eq:dressed}
\ee
and
\be
i P_{\mu\nu;\alpha'\beta'}(-i\Sigma^{\alpha'\beta';\mu'\nu'})i P_{\mu'\nu';\alpha\beta}=iP_{\mu\nu;\alpha\beta}\Sigma(s)  \,.
\ee

Using dimensional regularization and the $\overline{\rm MS}$ renormalization scheme, one obtains for the renormalized self-energy $\widetilde\Sigma\equiv\widetilde\Sigma_R +i\widetilde\Sigma_I$ in the second Riemann sheet (see App.~\ref{sec:appendix-scalar})
\begin{align}
\widetilde  \Sigma_R^{\rm II} &= \frac{\kappa^2}{(120 \pi)^2}\left[- 30s^2\lambda^5 \mathcal F(-\lambda) - 1155m^4 + 430m^2 s - 46s^2 \right. \nonumber \\
    &\left. + 15(30m^4-10m^2 s+s^2)\log(m^2/\mu^2)\right]  \,,  \label{eq:ReSigmaII-scalar} \\
\widetilde  \Sigma_I^{\rm II} &=- \frac{\kappa^2}{960 \pi } s ^2 \lambda^{5} \, \Theta(s - 4 m^2) \,,  \label{eq:ImSigmaII-scalar}
\end{align}
where $\lambda\equiv\sqrt{1-4m^2/s}$, and the function $\mathcal F$ is defined in (\ref{eq:functionF}). We will identify the renormalization scale with the cutoff $\mu=M_5$~\footnote{In the absence of any renormalization group improvement, the natural choice for the scale $\mu$ is the UV cutoff $M_5$.}.

\subsubsection{Seagull diagram}
In App.~\ref{sec:appendix-tadpole} we also have computed the contribution from the seagull diagram, coming from the $hh\phi\phi$ interaction. The result is $s$-independent and given by
\be
\widetilde \Sigma=-\frac{1}{64\pi^2}\left(7+6\log\frac{M_5^2}{m^2}  \right)\frac{m^4}{M_5^3} \,. \label{eq:Sigma_seagull}
\ee

\subsection{Fermions}

Let us here consider a Dirac fermion with mass $m$. Its Lagrangian is given by
\be
\mathcal L=\frac{i}{2}(\bar\psi \gamma^\mu\partial_\mu \psi-(\partial_\mu\bar\psi)\gamma^\mu\psi)-m\bar\psi\psi  \,,
\ee
and the energy-momentum tensor by
\be
T_{\mu\nu}=\frac{i}{4}\bar\psi(\gamma_\mu \overset{\leftrightarrow}{\partial_\nu}+\gamma_\nu \overset{\leftrightarrow}{\partial_\mu} )\psi-\eta_{\mu\nu}\mathcal L \,.
\label{eq:tensor-fermion}
\ee
The vertex for a graviton (with incoming momentum $p$) and one fermion (with incoming momentum $p_1$) and another fermion (with outgoing momentum $p_2$) is 
\be
i\Gamma_{\mu\nu}=-i\frac{\kappa}{4}\left[\gamma_{\mu}(p_1+p_2)_\nu+\gamma_\nu(p_1+p_2)_\mu -2\eta_{\mu\nu}(\slashed{p}_1+\slashed{p}_2-2m)\right] \,.
\label{eq:coupling-fermion}
\ee

Using dimensional regularization and the $\overline{\rm MS}$ renormalization scheme, the renormalized self-energy in the second Riemann sheet is given by (see App.~\ref{sec:appendix-fermion})
\begin{align}
\widetilde\Sigma_R^{\rm II}&= - \frac{\kappa^2}{7200\pi^2}\left[30 (8 m^2 + 3s) s\lambda^{3}\mathcal F(-\lambda) - 1635 m^4 - 190 m^2 s+ 108 s^2\right.\nonumber\\
&\left.+15(30m^4+10m^2 s - 3s^2)\log\frac{m^2}{\mu^2}\right] \,, \label{eq:ReSigmaII-fermion}\\
\widetilde\Sigma_I^{\rm II}&= -\frac{\kappa^2}{480\pi} (8 m^2 + 3s) s\lambda^{3}\,\Theta(s-4m^2)  \,. \label{eq:ImSigmaII-fermion}
\end{align}

\subsection{Massive gauge sector}

 Here we consider the case of massive (i.e.~$W^\pm,Z$)  gauge bosons in the Feynman gauge and the corresponding Faddeev-Popov ghosts.

\subsubsection{Gauge bosons}

We start with the Proca Lagrangian for a massive gauge boson with mass $m$,
\be
\mathcal L_{\rm Proca}=-\frac{1}{4}F_{\mu\nu}F^{\mu\nu}+\frac{1}{2}m^2 A_\mu A^\mu  \,,
\ee
and energy-momentum tensor
\be
T^{\mu\nu}=-F^{\mu\alpha}F^\nu_{\alpha}+\frac{1}{4}\eta^{\mu\nu}F_{\alpha\beta}F^{\alpha\beta}+m^2\left(A^\mu A^\nu-\frac{1}{2}\eta^{\mu\nu}A_\alpha A^\alpha\right)  \,.
\label{eq:tensor-massivegb}
\ee

The vertex for the interaction of a massive graviton $h^{\mu\nu}$ (with incoming momentum $p$) with gauge bosons $A^\sigma$  and $A^\rho$ (with incoming momenta $p_1$ and $p_2$) is
\begin{align}
i\Gamma^{\mu\nu,\rho\sigma}&=-i\kappa \left[C^{\mu\nu,\rho\sigma}+m^2D^{\mu\nu,\rho\sigma}  \right] \,, \label{eq:coupling-massivegb}\\
C^{\mu\nu,\rho\sigma}&=\frac{1}{2}\left[\eta^{\rho\sigma}(p_1^\mu p_2^\nu+p_1^\nu p_2^\mu)-\eta^{\mu\nu}(p_1^\rho p_2^\sigma+p_1^\sigma p_2^\rho)\right.\nonumber\\
&-(\eta^{\mu\rho}p_1^\sigma+\eta^{\mu\sigma}p_1^\rho)p_2^\nu-(\eta^{\nu\rho}p_1^\sigma+\eta^{\nu\sigma}p_1^\rho)p_2^\mu \label{eq:coupling-massive-gauge}\\
&\left.+(p_1\cdot p_2)(\eta^{\mu\rho}\eta^{\nu\sigma}+\eta^{\mu\sigma}\eta^{\nu\rho}+\eta^{\mu\nu}\eta^{\rho\sigma})\right] \,, \nonumber\\
D^{\mu\nu,\rho\sigma}&=\eta^{\mu\rho}\eta^{\nu\sigma}
+\eta^{\mu\sigma}\eta^{\nu\rho}-\eta^{\mu\nu}\eta^{\rho\sigma} \,,
\end{align}
which is manifestly symmetric under exchanges of $\mu\leftrightarrow\nu$ and $\rho\leftrightarrow\sigma$,
while the gauge boson propagator in the Feynman gauge is
$
P^{\rho\sigma}=-i\eta^{\rho\sigma}/(k^2-m^2)\,
$.

In the $\overline{\rm MS}$ scheme, the renormalized self-energy  in the second Riemann sheet is given by (see App.~\ref{sec:appendix-massivegb})
\begin{align}
\widetilde\Sigma_R^{\rm II}&= \frac{\kappa^2}{(120 \pi)^2} \Bigg[ 15 (60 m^4 + 25 m^2 s + 3 s^2) \log(m^2/\mu^2) - 2 (1695 m^4 + 455 m^2 s + 24 s^2)  \nonumber \\
      & - 30 (98 m^4 + 31 m^2 s + 3 s^2) \lambda(s) \mathcal F(-\lambda )\Bigg] \,,  \label{eq:ReSigmaII-massivegb} \\
\widetilde\Sigma_I^{\rm II}&= -\frac{\kappa^2}{960\pi}(98 m^4 + 31 m^2 s + 3 s^2) \lambda(s)\, \Theta(s-4m^2)  \label{eq:ImSigmaII-massivegb}\,.
\end{align}

\subsubsection{Faddev-Popov ghosts}

The energy-momentum tensor and vertex to the massive graviton for Faddev-Popov (FP) fields $c$ and $\bar c$ (with incoming momenta $p_1$ and $p_2$) are, in the Feynman gauge (for which its mass is equal to the gauge boson mass $m_c=m$), given by
\be
T^{\mu\nu}=\partial^\mu\bar c\partial^\nu c +\partial^\nu\bar c\partial^\mu c-\eta^{\mu\nu}(\partial_\alpha \bar c\partial^\alpha c+m^2 \bar c c)  \,,
\label{eq:tensor-massiveFP}
\ee
and
\be
i\Gamma^{\mu\nu}(p_1,p_2)=i\kappa\left[p_1^\mu p_2^\nu+p_1^\nu p_2^\mu-\eta^{\mu\nu}(p_1\cdot p_2-m^2) \right]  \,.
\label{eq:coupling-massiveFP}
\ee

The result for the $\overline{\rm MS}$ renormalized self-energy in the second Riemann sheet is given by (see App.~\ref{sec:appendix-massiveFP})
\begin{align}
\widetilde\Sigma_R^{\rm II}&=\frac{\kappa^2}{7200\pi^2}\left[30s^2\lambda^{5}\mathcal F(-\lambda)+1155m^4-430m^2 s+46s^2\right. \label{eq:ReSigmaII-massiveFP} \nonumber\\
&\left.-15(30m^4-10m^2 s+s^2)\log(m^2/\mu^2)\right]  \,, \\
\widetilde\Sigma_I^{\rm II}&=\frac{\kappa^2}{480\pi}s^2\lambda^{5}\,\Theta(s-4m^2) \label{eq:ImSigmaII-massiveFP} \,.
\end{align}

\subsection{Massless gauge sector}
 Here we consider the case of massless (i.e.~$\gamma, g^A$) gauge bosons and the corresponding Faddeev-Popov ghosts.
 
 \subsubsection{Gauge bosons}
   The interaction of massive gravitons with massless gauge bosons is given by Eq.~(\ref{eq:coupling-massive-gauge}) with $m=0$. The final result for the renormalized self-energy in the second Riemann sheet is given by (see App.~\ref{sec:appendix-masslessgb})
  \begin{align}
  \widetilde\Sigma_R^{\rm II} &=  -\frac{\kappa^2}{4800\pi^2}s^2\left[16-15\log(s/\mu^2)  \right]   \,, \label{eq:ReSigmaII-masslessgb} \\
  \widetilde\Sigma_I^{\rm II} &=- \frac{\kappa^2}{320 \pi } s^2 \,.  \label{eq:ImSigmaII-masslessgb}
 \end{align}
 These results apply both to the photon and to the $N_c^2-1$ gluons.

\subsubsection{Faddeev-Popov ghosts}
For FP ghosts corresponding to massless gauge fields (i.e.~$\gamma$ and $g^A$) the result in the second Riemann sheet is (see App.~\ref{sec:appendix-masslessFP})
\begin{align}
\widetilde\Sigma_R^{\rm II}  &=  \frac{\kappa^2}{7200\pi^2} s^2[46-15\log(s/\mu^2)] \,, \label{eq:ReSigmaII-masslessFP} \\
\widetilde\Sigma_I^{\rm II}  &= \frac{\kappa^2}{480\pi} s^2   \,. \label{eq:ImSigmaII-masslessFP}
\end{align}

 \section{Long-lived massive graviton dark matter}
\label{sec:longlived_DM}

 We will consider the case of massive gravitons propagating in the warped 5D space defined in section~\ref{sec:model}.

 \subsection{The isolated resonance: The massive graviton}

 The brane-to-brane propagator of the massive gravitons, either in the subspace $\mathcal M^+$ or in $\mathcal M^-$, was computed in section~\ref{subsec:graviton_propagator} (see also \cite{Fichet:2026nct}), and it is given by
 \be
 G_h(s)=\frac{-1}{m_g+\sqrt{m_g^2-s}}  \,.
 \ee 
In this expression for the propagator any graviton zero mode has been subtracted. 

Radiative corrections generate an isolated pole in the second Riemann sheet of the complex $s$-plane at $s_p=m_p^2-i m_p \Gamma_p$ as the zeros of the equation
 \be
 D_h(s_p)-\Sigma_R(s_p)-i \Sigma_I(s_p)=0,\quad D_h(s)=-m_g-\sqrt{m_g^2-s}  \,,
 \ee
 where $\Sigma_R\equiv\textrm{Re} \, \Sigma$ and $\Sigma_I\equiv\textrm{Im} \, \Sigma$, and $m_p<m_g$.
 
 We will consider the contribution of all states localized on the brane, and simplify the analysis by assuming that the heaviest state contributing to the graviton self-energy has a mass $m\gg m_g$. We will consider such state as a heavy scalar $s=\phi$, e.g.~the inflaton responsible for cosmological inflation (see e.g.~Ref.~\cite{Mishra:2025ofh}). In that case the full value of $\Sigma_R$ is dominated by this heaviest state, and in fact $\Sigma_R(s)\simeq\Sigma_R^s$ is a constant given by (see App.~\ref{sec:appendix-scalar})
 \be
 \Sigma_R^s \simeq -\frac{1}{32\pi^2}\left(5+4\log\frac{M_5^2}{m^2}  \right)\frac{m^4}{M_5^3},\quad \Sigma_I^s = 0  \,,
 \label{eq:SigmaRs}
 \ee
 where $m$ is the inflaton mass, while the lighter SM states coupled to the graviton can contribute to $\Sigma_I$.
 
 Necessary and sufficient conditions for the existence of a nontachyonic pole are
 \be
 0\leq -m_g-\Sigma_R\leq m_g,\quad \Sigma_I(m_p^2) < 0  \,,
 \label{eq:conditions}
 \ee
 which can be written as
 \be
 \Sigma_R=-a\,m_g \,,\quad\textrm{with}\quad a\in [1,2] \quad \Longrightarrow\quad m_p\in[m_g,0] \,,
 \label{eq:condition2}
 \ee
 where the case $a=1$ corresponds to $m_p=m_g$, and $a=2$ to $m_p=0$. 
 
 The condition (\ref{eq:condition2}) yields for the mass $m$ the value
 \be
 \frac{m(a)}{M_5}=\left(\frac{e^{5/2}x}{-\mathcal W[-x]}  \right)^{1/4},\quad \textrm{with}\quad x=\frac{16\pi^2 a\,m_g}{e^{5/2}M_5}  \,,
 \label{eq:maM5}
 \ee
 where $\mathcal W$ is the principal branch of the Lambert function, so that we can trade the parameter $m$ by $a$. We will postpone more details about the solution till the end of the section.

The location of the pole is then given by
 \begin{align}
 m_p&=\sqrt{-2 m_g \Sigma_R-\Sigma_R^2}=\sqrt{a(2-a)}\, m_g \,, \nonumber\\
 \Gamma_p&=\frac{2(m_g+\Sigma_R)}{m_p}\Sigma_I(m_p^2)=-2\sqrt{\frac{(a-1)^2}{a(2-a)}}\Sigma_I(m_p^2)  \,.
 \end{align}

\begin{figure}[htb]
\begin{center}  
 \includegraphics[width=0.48\textwidth]{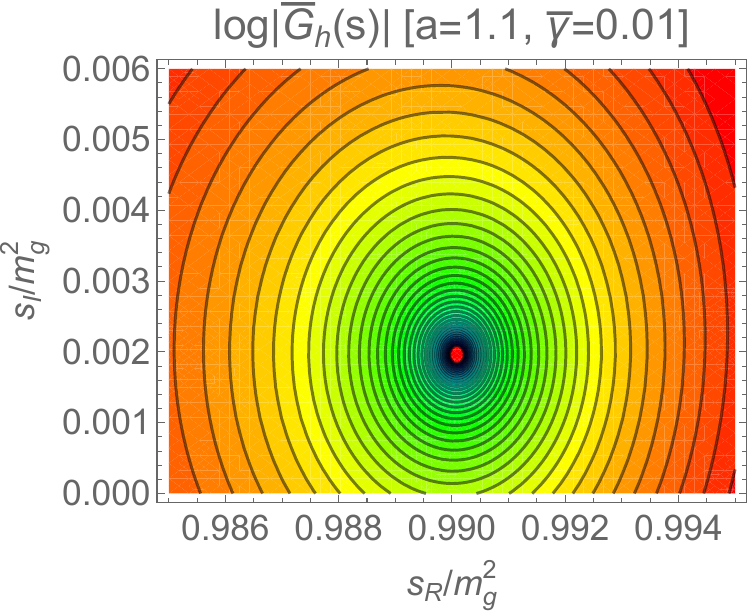}     \includegraphics[width=0.51\textwidth]{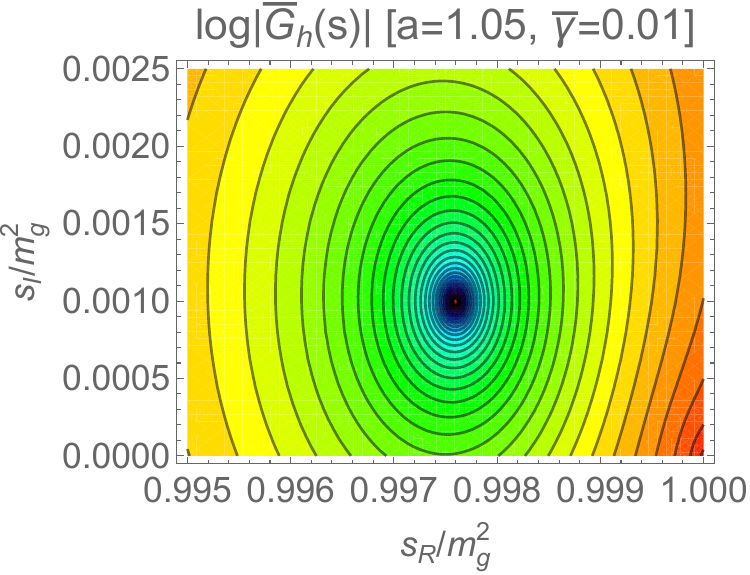}  
\end{center}
     \caption{\it Left panel: contour lines of $|\overline G_h(s)|$ in the plane $s=s_R-i s_I$ for $a=1.1$ and $\bar \gamma \equiv m_g^3 \gamma =0.01$. Right panel: the same for $a=1.05$. At the pole: $s_R=m_p^2$ and $s_I=m_p\Gamma_p$. The red (blue) regions indicate lower (higher) values of $|\overline G_h(s)|$.}
    \label{fig:pole}
 \end{figure}

  To check the analytical approximation for the solution of the pole equation, we have considered a particular simple model where the real part $\Sigma_R$ is led by the heavy scalar $\phi$ so that we have already incorporated its effect on the resummed propagator
  \begin{equation}
   \bar G_h(s)\simeq-\left(\sqrt{m_g^2-s}-(a-1)m_g  +i \Sigma_I\right)^{-1} \,,
\end{equation}
  and the imaginary part of the self-energy is parametrized as
  \begin{equation}
    \Sigma_I(s)=- \gamma s^2 \,,
  \end{equation}
  where $\gamma$ is an arbitrary constant with dimension $-3$. The
  pole mass and width are given by $m_p(a)=\sqrt{a(2-a)} \, m_g$, so
  that $\Sigma_I(m_p^2) = - a^2(2-a)^2 \gamma m_g^4$, and
  $\Gamma_p(a)=2 a^{3/2}(2-a)^{3/2}(a-1) \gamma m_g^4$. Contour lines
  of $|\overline G_h|$, where $\overline
  G_h=[G_h^{-1}(s) - \Sigma_R - i\Sigma_I(s)]^{-1}$, in the plane $s=s_R-i s_I$ are
  shown in Fig.~\ref{fig:pole} for the cases $a=1.1$ (left panel) and
  $a=1.05$ (right panel) and for $\bar\gamma \equiv m_g^3 \gamma =
  0.01$.
The analytical approximation of masses and widths are: (i) for $a=1.1$: $m_p^2\simeq 0.990\, m_g^2$, $\Gamma_p\simeq 0.002\, m_g$; and, (ii) for $a=1.05$: $m_p^2\simeq 0.998\, m_g^2$, $\Gamma_p\simeq 0.001\, m_g$. These values are well reproduced by the poles shown in the left and right panels of Fig.~\ref{fig:pole}.

 The residue of the pole is given by
 \be
 R(m_p)=\frac{1}{D_h^\prime(s)-\Sigma^\prime_R(s)|_{m_p^2}}= -2(m_g+\Sigma_R^s)+ \dots\simeq2(a-1)m_g \,,
 \ee
  and the coupling of the isolated resonance $\chi_{\mu\nu}$ to the energy-momentum tensor of matter in the brane is given by
  \be
  \kappa_\chi\equiv\frac{1}{M_5}\left(\frac{R(m_p)}{M_5} \right)^{1/2}\simeq \frac{\sqrt{3(a-1)}}{M_4} \equiv \frac{\lambda_\chi}{M_4} \,,
  \ee 
  where $\lambda_\chi$ is the Wilson coefficient and the 4D Planck scale $M_4$ the cutoff, defined in the interval $\lambda_\chi\in[0,\sqrt{3}]$.
  
  In fact we can use the Wilson coefficient $\lambda_\chi$ as the free parameter, subject to the constraint $0<\lambda_\chi<\sqrt{3}$, and the pole mass and width of the isolated resonance are given by
  \begin{align}
    m_p&=\sqrt{1-\frac{\lambda_\chi^4}{9}}m_g \,, \qquad
    \Gamma_p=-2 \frac{\frac{\lambda_\chi^2}{3}}{\sqrt{1-\frac{\lambda_\chi^4}{9}}}\Sigma_I(m_p^2)\,.
  \end{align}
  In particular for $\lambda_\chi\ll 1$ we have that $m_p\simeq m_g$ while $\Gamma_p$ is strongly suppressed by $\sim \lambda_\chi^2$, a property that will be used later on.
  
  We will now conclude this section with some comments about the solution to the pole equation (\ref{eq:condition2}), which can be written as
  \be
  \Sigma_R+m_g=-(a-1)m_g=-\frac{\lambda_\chi^2}{3}m_g\,.
  \label{eq:condition3}
  \ee
  Assuming that $\Sigma_R$ is dominated by the heavy scalar $\Sigma_R\simeq \Sigma^s_R$, Eq.~(\ref{eq:condition3}) can be satisfied by tuning the renormalization scale, here identified with the 5D cutoff $M_5$, as
  \be
  M_5=e^{-5/8}m+4\pi^2 e^{-5/2} a \, m_g+\cdots \,,
  \label{eq:M5mexpansion}
  \ee
  or equivalently,
  \be
  m_g=\frac{3}{2}\frac{(e^{-5/8}m)^3}{M_4^2}+27\pi^2 e^{-5/2} a \frac{(e^{-5/8}m)^5}{M_4^4}+\cdots \,,
  \ee
  where the above expansions are assuming that $m_g\ll m\ll M_4$. In fact the first term in (\ref{eq:M5mexpansion}) makes the radiative correction $\Sigma^s_R$ to vanish, while the second one is tuned to find the required solution. 
  \begin{figure}[htb]
\begin{center}  
  \includegraphics[width=0.45\textwidth]{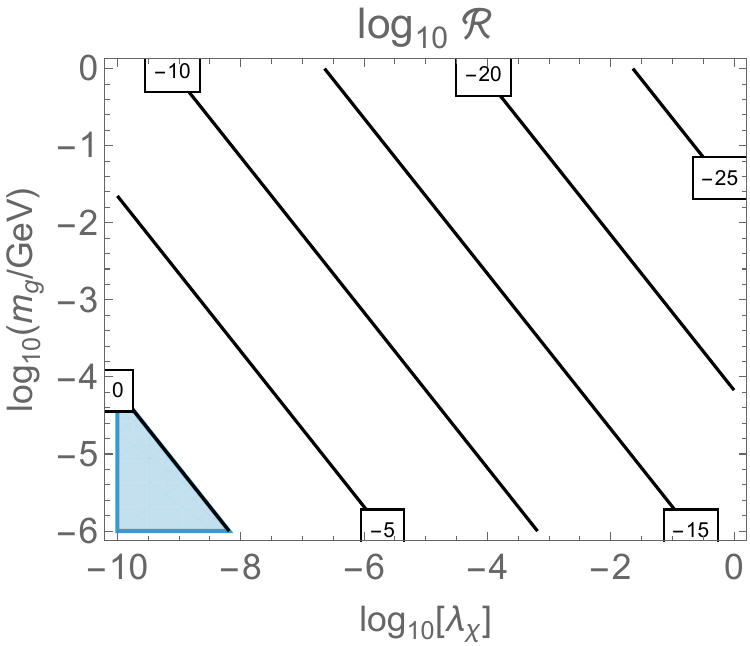}     
\end{center}
     \caption{\it Contour lines of $\mathcal R$ in the plane $(\lambda_\chi,m_g)$. The shadowed region does not satisfy condition (\ref{eq:condition4}).}
     \label{fig:consistency}
 \end{figure}
  \newline\indent
  However the total self energy is given by $\Sigma_R=\Sigma_R^s+\Sigma_R^{\rm SM}$ so that once we have fixed the value of $M_5$, to not disturb the solution  to (\ref{eq:condition3}) the condition
  \be
 \mathcal R\equiv \frac{3}{\lambda_\chi^2}\, \frac{\Sigma_R^{\rm SM}}{m_g}\ll 1
 \label{eq:condition4}
  \ee
  must be satisfied.
  The value of $\Sigma_R^{\rm SM}$ is dominated by the heaviest SM particle, i.e.~the top quark, for which
  \be
  \Sigma_R^t\simeq \frac{1}{32\pi^2}\left(3+2\log\frac{M_5^2}{m_t^2} \right)\frac{m_t^4}{M_5^3}\,.
  \label{eq:SigmaRtop}
  \ee
  In fact, as $M_5$ is a function of $m_g$, condition (\ref{eq:condition4}) provides an allowed region in the plane $(\lambda_\chi,m_g)$, where $\mathcal R\ll 1$, as shown in the plot of Fig.~\ref{fig:consistency}.

  \subsection{The isolated resonance as dark matter}
  
  Depending on the value of $\Gamma_p$ the resonance can decay on cosmological times, or with a lifetime larger than the age of the universe. In the latter case the resonance can be a candidate  to dark matter. In the former case the resonance will decay and could eventually be detected at future accelerators, or by its indirect effects. Both possibilities depend on the value of $\Gamma_p$ which in turns depends on the value of $\Sigma_I(m_p)$. 

\begin{figure}[htb]
\begin{center}  
  \includegraphics[width=0.45\textwidth]{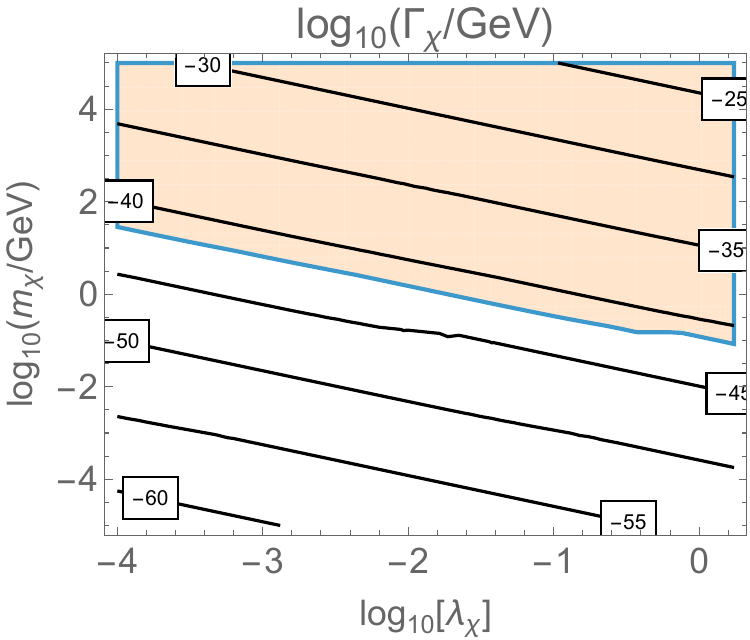}     \includegraphics[width=0.45\textwidth]{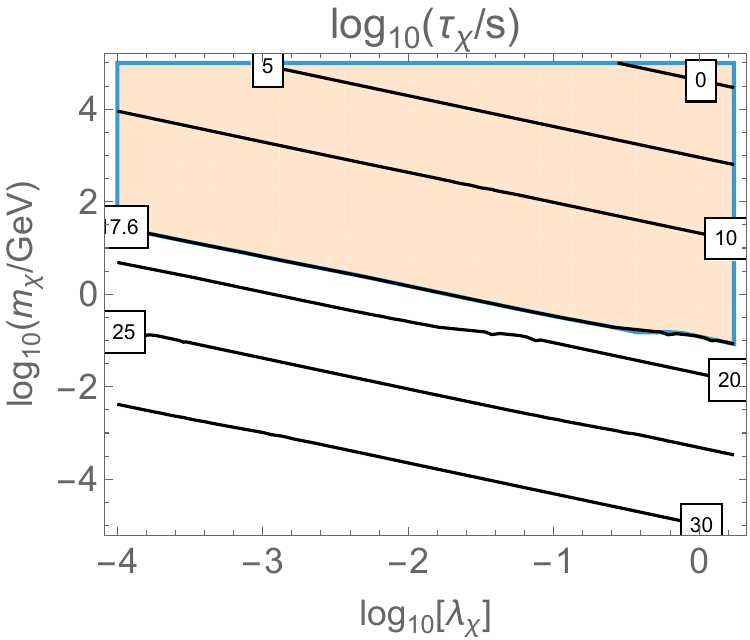}
\end{center}
     \caption{\it Left panel: contour lines of $\Gamma_p$ in the plane $(\lambda_\chi,m_p)$. Right panel: the same for the lifetime $\tau_p$. In the white areas $\tau_p>\tau_U$, where $\tau_U$ is the present age of the universe.}
     \label{fig:plotGammap}
 \end{figure}

   As for the imaginary part of $\Sigma$, it is contributed by the SM fields $\varphi$, such that $m_p^2>4m_\varphi^2$, as~\footnote{Note that every contribution corresponding to the SM field $\varphi$, with mass $m_\varphi$ contains a step function $\theta(s-4m_\varphi^2)$ and then vanishes for $s<4m_\varphi^2$.}
   \be
   \Sigma_I^{\rm SM}=\Sigma_I^s(m_H)+\sum_{V=W,Z}N_V[\Sigma_I^s(m_V)+\Sigma_I^v(m_V)]+ 9\Sigma^v_I(0)+\sum_f N_c\Sigma^f_I(m_f)
      \ee
  evaluated at $s=m_p^2$,  where $N_W=2$, $N_Z=1$, and the contribution from FP fields is already included in $\Sigma_I^v$. The factor $N_c$ is the number of colors, i.e.~$N_c=3\,(1)$ for quarks (leptons).
 The width $\Gamma_p$ is shown in the left panel of Fig.~\ref{fig:plotGammap} while the corresponding lifetime is in the right panel of Fig.~\ref{fig:plotGammap}. The white areas correspond to the region where the lifetime is larger than the present age of the universe, so that the particle is stable on cosmological times. In the shadowed areas the state is unstable but it decays always outside the detector, so that it can only provide a missing energy signal.   

  \subsection{The freeze-in generation of dark matter}
  
  The isolated resonance $\chi_{\mu\nu}$ then corresponds to a FIMP, a massive graviton that in the unshaded region of Fig.~\ref{fig:plotGammap} has a lifetime larger than the age of the universe $\tau_p>\tau_U$ and is thus a candidate for the dark matter of the universe. Assuming that the inflaton is mainly coupled to the SM fields (as it is only gravitationally coupled to the massive graviton), after inflation, at the reheating temperature $T_R$, the SM is a plasma in thermal equilibrium while the FIMP is out of equilibrium  with zero density. Its energy density must be produced by the freeze-in mechanism.
  
\begin{figure}[htb]
\begin{center}  
  \includegraphics[width=0.45\textwidth]{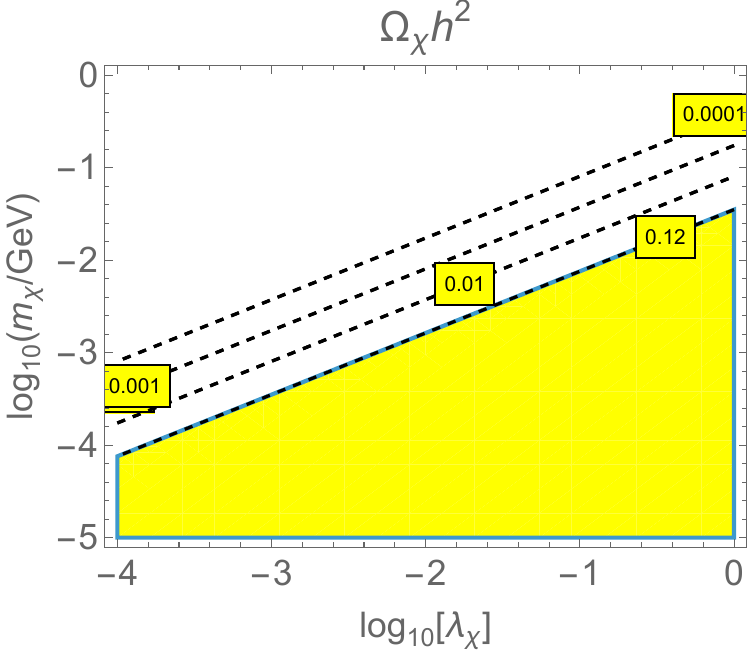}     \includegraphics[width=0.45\textwidth]{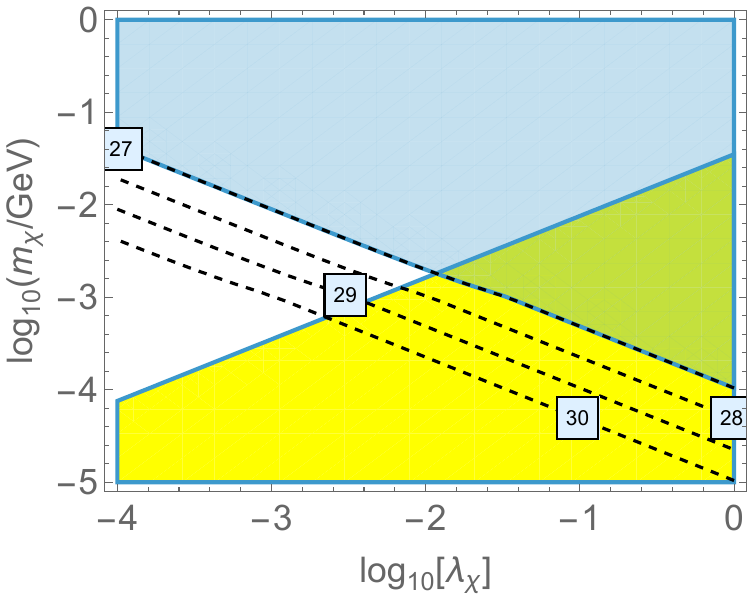}
\end{center}
     \caption{\it Left panel: contour (solid) lines of $\Omega_\chi h^2$ in the plane $(\lambda_\chi,m_p)$. Right panel: the upper (blue) shadowed region corresponds to $\tau_{\chi}<10^{27}$s. Contour (dashed) lines corresponding to other values of $\log_{10}(\tau_\chi/$s) are also shown. The lower (yellow) shadowed regions in both panels correspond to $\Omega_\chi h^2>0.12$, and are forbidden by overclosure of the universe. In the white region of the right panel the massive graviton can be (a component of) long-lived dark matter.}
     \label{fig:Omegachi}
 \end{figure}

  The process $\varphi\varphi\to\chi$ is too suppressed to yield a sizable energy density given that the inverse process, the decay $\chi\to\varphi\varphi$, is required to be extremely small for the massive graviton to not decay on cosmological times. The freeze-in production by the 2 to 2 processes $\varphi_1\varphi_2\to\varphi_3\chi$ (where $\varphi_{1,2,3}$ are SM particles) was studied in great detail in Ref.~\cite{Cai:2021nmk}. It was found that the freeze-in is dominated by the channels involving the QCD coupling $q\bar q\to g\chi$ and $qg\to q\chi$, where $q=c,b$, are heavy enough quarks to trigger the effect but light enough to be in thermal equilibrium below the critical temperature $T_c\simeq 160$ GeV of electroweak symmetry breaking~\cite{Quiros:1999jp}, and $g$ are the gluons. Moreover the resulting energy density is IR dominated (the temperature integral is dominated by temperatures $T\sim T_c$), and thus insensitive to the reheating temperature (but sensitive to the value of the involved quark masses above $T_{\rm QCD}\simeq 150$ MeV), while the UV contribution is subleading. The final yield gives~\cite{Cai:2021nmk}
  \be
  \Omega_\chi h^2\simeq 5.2\times 10^{-6}\lambda_\chi^2\, \left(\frac{\textrm{GeV}}{m_\chi}\right)^3  \,.
  \ee
  We show in the left panel of Fig.~\ref{fig:Omegachi} contour lines of $\Omega_\chi h^2$ in the plane $(\lambda_\chi,m_p)$. The shaded (yellow) region is excluded by overclosure of the universe.

There are stringent bounds on decaying DM from indirect production and cosmological observations. In particular DM can be constrained from observations of the galactic and extra-galactic diffuse x-ray and gamma-ray backgrounds yielding upper bounds on $\tau_{\chi}$~\cite{Essig:2013goa}. Moreover, there are also cosmological constraints on exotic injection of electromagnetic energy and its effect on the CMB power spectra~\cite{Poulin:2016anj,Slatyer:2016qyl}, and effects of the decay channels $e^+ e^-$ and $\gamma\gamma$ in the 21-cm power spectrum~\cite{Facchinetti:2023slb,Sun:2023acy}. All of them fix a global lower limit on the DM lifetime, that we take as $\tau_\chi\gtrsim 10^{27}$ s. In the right panel of Fig.~\ref{fig:Omegachi}, the upper shadowed (blue) region is where $\tau_{\chi}<10^{27}$ s, and the lower shadowed (yellow) region where $\Omega_\chi h^2>0.12$. 
In the white area the massive graviton can be long-lived  dark matter, and at its lower border it saturates the dark matter abundance today. This provides upper bounds on $m_\chi$ and $\lambda_\chi$, given by $m_\chi\lesssim 2$ MeV and $\lambda_\chi\lesssim 0.01$. Contour lines corresponding to $\tau_\chi=10^{28}-10^{30}\,$s are also shown in view of future cosmological bounds on $\tau_\chi$.

 \subsection{Final comments}
 The range for the isolated resonance, and the inflaton mass $m$, to be identified as DM is then
$
m_\chi\lesssim 2 \textrm{ MeV} \; \textrm{and} \;  m\lesssim 4\times 10^{11} \textrm{ GeV}\,,
$
while there is no theoretical lower bound, although Lyman-$\alpha$ forest data~\cite{Viel:2013fqw,Viel:2007mv}, as well as updated Milky Way satellite counts, and strong gravitational lensing, put experimental lower bounds on warm dark matter (WDM) mass around $m_\chi\gtrsim 20$ keV~\cite{DEramo:2025jsb}, which corresponds to $m\gtrsim 8\times 10^{10}$ GeV. This yields the allowed bands in the fundamental parameters of our theory that are provided in table~\ref{tab:summary}.
\begin{table}[htb]
  \centering
  \begin{tabular}{|l||c||}
    \hline
    \multicolumn{2}{|c|}{\bf Fundamental parameters} \\
    \hline
    \hline
     {\bf DM mass} ( $\simeq$ {\bf mass gap})  &   $20\textrm{ keV}\lesssim m_\chi\lesssim 2\textrm{ MeV}$  \\
   \hline
        {\bf 5D Planck mass}  &  $4 \times 10^{10} \textrm{ GeV} \lesssim M_5 \lesssim 2 \times 10^{11}\textrm{ GeV}$  \\
    \hline
 {\bf Inflaton mass}    &  $8\times 10^{10} \textrm{ GeV}\lesssim m\lesssim 4\times 10^{11}\textrm{ GeV}$   \\
     \hline 
  \end{tabular}
\caption{\it Summary of the range of values of the fundamental parameters of the theory.}
\label{tab:summary}
\end{table}

This sub-MeV dark matter could be thought to be helpful in solving some of the typical problems of the cold dark matter (CDM) scenario, fitting small-scale observations~\cite{DES:2020fxi}. The free-streaming of sub-MeV dark matter can erase small-scale density fluctuations and thus help to solve the missing satellites problem: this is the so-called warm dark matter scenario. However for WDM the free streaming wavelength $\lambda_{\rm FS}\sim 0.3 \,(1\textrm{ keV}/m_\chi)^{4/3}$ Mpc is only effective for $m_\chi\sim 1$ keV, for which $\lambda_{\rm FS}\sim 0.3$ Mpc (the scale of dwarf galaxies), while for   
$m_\chi\sim 10$ keV the free streaming length is $\lambda_{\rm FS}\sim 0.006$ Mpc, in tension with observations~\cite{DES:2020fxi}, and the difference with respect to CDM is indistinguishable. Thus the range of parameters in table~\ref{tab:summary} is in the CDM ballpark.

To conclude this section we want to comment on some experimental prospects for the proposed long-lived massive graviton as a dark matter candidate. A massive graviton with only gravitational-strength interactions is extremely
difficult to probe in conventional laboratory dark-matter experiments.  As its
coupling to the Standard Model is suppressed by a very high scale, nuclear or
electron recoil rates are negligible, and collider production of the dark-matter
state itself is generally unobservable.  The most promising probes are therefore
often astrophysical and cosmological.
\newline\indent
As the resulting dark matter can
naturally lie in the 20 keV--2 MeV range, where the lower edge is associated with warm-dark-matter constraints from
Lyman-$\alpha$, Milky Way satellites and strong-lensing observations, the model is better probed by decaying dark matter. For the decay
$ \chi\rightarrow\gamma\gamma$, the line energy is
$ E_\gamma=\frac{m_\chi}{2}$.
Therefore the $0.02$--$2$ MeV mass range maps onto photon energies of roughly
$10$ keV--$1$ MeV.
This is experimentally attractive because it lies in the poorly explored
hard-x-ray/MeV transition region.  The International Gamma-Ray Astrophysics Laboratory Spectrograph (INTEGRAL/SPI)~\cite{Attie:2003qf,Knodlseder:2003sv,Weidenspointner:2006nua} has already constrained narrow
decaying-dark-matter lines corresponding to dark-matter masses of roughly
$40$ keV--$14$ MeV, with limits that are typically around
$10^{27}$--$10^{28}$ s depending on mass, channel and analysis.  Systematic
uncertainties from instrumental backgrounds and the Galactic dark-matter
profile are important.
The upcoming NASA $\gamma$-ray Compton Spectrometer and Imager (COSI) space telescope~\cite{Tomsick:2023aue}, scheduled for launch in 2027, with sensitivity in the
roughly $0.2$--$5$ MeV photon-energy range, is especially well matched to the
upper part of the isolated-graviton mass window.  Proposed MeV missions such
as the All-Sky Medium Energy Gamma-Ray Observatory eXplorer (AMEGO-X)~\cite{Fleischhack:2023ube} would extend coverage toward lower photon energies and broader
energy ranges.
\newline\indent
For
$
 m_\chi>2m_e\simeq1.022\ {\rm MeV},
$
the decay
$
 \chi\rightarrow e^+e^-
$
becomes kinematically allowed.
This creates an additional observational channel.  The injected positrons
can lose energy and eventually annihilate, producing 511 keV photons and a
continuum component.  The resulting signal depends not only on the decay
rate but also on positron propagation, energy loss, magnetic fields and the
spatial distribution of dark matter.  Therefore, a realistic 511 keV forecast
requires a Galactic propagation calculation rather than simply equating the
decay rate with the observed line flux.

 \section{The connection with unparticles}
\label{sec:unparticles}
 
 The (scalar part of the) 5D brane-to-brane inverse propagator is given by
 \be
 D_h=-m_g-\sqrt{m_g^2-s}  \,,
 \ee
 to be compared with the one for gapped unparticles with scaling dimension $d_U$ where
 \be
 D_{\rm un}\propto -(m_g^2-s)^{2-d_U}\,.
 \ee
 
 While in our 5D construction the mass gap $m_g$ has its origin in the brane potential of the stabilizing field leading to the VEV $\bar v_b$, as $m_g=(3/2) k e^{\bar v_b}$, and comparison between both propagators suggests that in our case $d_U=3/2$, as implied by the 5D theory, identification between both propagators  fails because of the constant term in $D_h$. We will see now how to recover the unparticle structure out of our 5D propagator. 
 
 We have seen in previous sections how interactions of the localized matter with the bulk gravitons lead to renormalization of the self-energy propagator as $D_h\to D_h-\Sigma(s)$, and the appearance of an isolated resonance. We have considered the contribution from the SM fields and a scalar (that we can identify with the inflaton) $\phi$, with mass $m$ much larger than all SM particles $m\gg m_\varphi$. We will mainly concentrate on the contribution from $\Sigma_\phi$ and will consider the region $s\gg m^2$, where $\phi$ propagates, and the region $s\ll m^2$ where $\phi$ is decoupled.
 
 \subsection{Region $s\gg m^2$}
 
 From the loop diagram contributions to the self-energy where both $\phi$ and all SM particles are propagating, the propagator $G_h$ becomes
 \be
 G_h^{(1)}=\frac{-1}{m_g+\sqrt{m_g^2-s}+\Sigma_\phi(s)+\Sigma_{\rm SM}(s)} \,,
 \ee
 where $\Sigma_\phi(s)$ is given by the sum of Eqs.~(\ref{eq:ReSigmaII-scalar}), (\ref{eq:ImSigmaII-scalar}) and (\ref{eq:Sigma_seagull}), while $\Sigma_{\rm SM}(s)$ is the contribution from all the SM particles.
 
 \subsection{Region $m_{\rm SM}^2 \ll s \ll m^2$}
 
 In this region the SM keeps propagating while $\phi$ decouples and $\Sigma_\phi(s)$ becomes a constant given by (\ref{eq:SigmaRs}). For the limiting case $a=1$, where the pole mass is $m_p= m_g$ the one-loop propagator is given by
 \be
 G_{\rm un}^{(1)}(s)=\frac{-1}{\sqrt{m_g^2-s}+\Sigma_R^{\rm SM}(s)+i\Sigma_I^{\rm SM}(s)} \,,
 \label{eq:unparticle}
 \ee
 which is the radiatively corrected propagator for an unparticle with dimension $d_U=3/2$ in the presence of the SM field interactions.
 
  In the case $a\lesssim 1$ the propagator has an extra contribution coming from $\phi$, a threshold effect, given by
 \be
  G_h^{(1)}(s)=\frac{-1}{\sqrt{m_g^2-s}-(a-1)m_g+\Sigma_R^{\rm SM}(s)+i\Sigma_I^{\rm SM}(s)} =\sum_{n=0}^\infty (-1)^{n}\delta^n (G_{\rm un}^{(1)})^{n+1} \,,
  \ee  
  where $\delta\equiv (\lambda_\chi^2/3)m_g$, which corresponds to an arbitrary number of $\delta$-insertions on the unparticle propagator.

 \subsection{Region $s< m_{\rm SM}^2$}
 
 In this region the SM fields $\varphi$ will also decouple when $s\ll m_\varphi^2$. Then the corresponding self-energy $\Sigma_\varphi$ will also contribute as a constant threshold. The SM effect on the pole location will then be dominated by the SM field with the highest mass, in particular the top mass $t$ with mass $m_t\simeq 173$ GeV, giving the threshold contribution of Eq.~(\ref{eq:SigmaRtop}) so that the location of the pole will be provided by the competition between the terms $ m_g$ and $\Sigma_R^{t}$. As $\mathcal R \ll 1$,  by many orders of magnitude (see Fig.~\ref{fig:consistency}), where $\mathcal R$ is given by Eq.~(\ref{eq:condition4}), the pole location $m_p$ is fixed by the contribution from the field $\phi$. We can then consider from here on, and from all practical purposes, that $G_h^{(1)}\simeq G_{\rm un}^{(1)}$.
 
 On the other hand the width of the resonance $\Gamma_\chi$ is fixed by the SM contribution $\Sigma_I^{\rm SM}(m_g^2)$, as in the case of the pure unparticle propagator (\ref{eq:unparticle}), and
 \be
  G_h^{(1)}(s)\simeq \frac{-1}{\sqrt{m_g^2-s}+i\Sigma_I^{\rm SM}(m_g^2)}  \,,
 \ee
 where only SM states $\varphi$ with masses $m_\varphi<m_g/2$ will contribute.

\section{The fluid production}
 \label{sec:fluid}

\begin{figure}[htb]
\begin{center}  
  \includegraphics[width=0.45\textwidth]{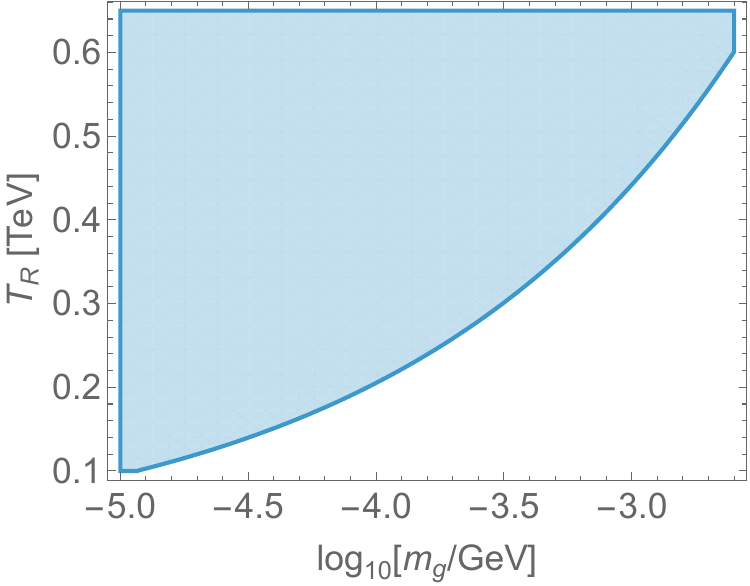} 
\end{center}
     \caption{\it Shadowed (unshadowed) region corresponds to $\rho_{\rm fluid}/s>0.44$ eV ($\rho_{\rm fluid}/s<0.44$ eV).}
     \label{fig:TRmg1}
 \end{figure}

As we have already seen in Ref.~\cite{Fichet:2026nct}, after cosmological inflation and reheating there is a fluid freeze-in production by the processes $\varphi\varphi\to h_{\mu\nu}$,  based on the leakage of gravitons from the brane into the bulk with an energy density rate 
 \be
 \Delta_h=-c\, T^8/M_5^3,\quad c = \frac{2C}{5\pi^4} \Gamma(7/2)\zeta(7/2)\Gamma(9/2)\zeta(9/2)  \,, \label{eq:Delta_h}
 \ee
where again all the SM fields $\varphi$ are contributing for values of the reheat temperature $T_R\gg m_\varphi$, while the inflaton does not if $T_R\ll m$, and then $C\simeq 20.78$ so that $c\simeq 3.92$. By using the explicit results of section~\ref{subsec:fluid}, the fluid energy density is thus given by
\be
\rho_{\rm fluid}=\frac{3\sqrt{5/2} \, c}{\sqrt{g_{\textrm{eff}}}\pi}\frac{T_R^3}{m_g M_4}T^3 \,,
\ee
which scales as matter, where $T$ is the SM (the photon) temperature. It is now common to compute the redshift invariant quantity $\rho/s$, where $s$ is the entropy density. In our case this yields
\be
\frac{\rho_{\rm fluid}}{s}=\frac{135}{2\pi^3}\frac{\sqrt{5/2}\, c}{g_{\textrm{eff}}^{3/2}} \frac{T_R^3}{m_g M_4}  \,,
\ee
where $g_{\textrm{eff}}$ is the number of relativistic degrees of freedom  in the thermal bath at freeze-in. As we can see the production is controlled by the reheating temperature $T_R$ which shows that the freeze-in production is UV dominated~\cite{Elahi:2014fsa}~\footnote{For nonstandard cosmologies, as  e.g.~in bigravity models, the UV freeze-in mechanism was worked out in Ref.~\cite{Bernal:2019mhf}.}. Assuming there is no entropy injection after freeze-in~\footnote{In fact for temperatures below $T_R$ the graviton leakage rate drops off very fast, as $T^8$.}, the observed dark matter abundance would imply that $\rho/s\leq 0.44$ eV. The condition $\rho_{\rm fluid}/s>0.44$ eV is forbidden by overclosure of the universe, and is shown in the shadowed region of Fig.~\ref{fig:TRmg1}, where only the interval $10 \textrm{ keV}<m_g<2 \textrm{ MeV}$ is shown. As is clear from table~\ref{tab:summary}, for larger values of~$m_g$ only the holographic fluid is a candidate to dark matter.
 Notice that along the border of the allowed (white) region $\rho_{\rm fluid}/s\simeq 0.44$ eV, while the value of $\rho_{\rm fluid}/s$ decreases very fast for smaller values of $T_R$ as it behaves as $T_R^3$. Along the border the holographic fluid can be dark matter (or at least a component thereof), while in the white area the isolated massive graviton we have discussed in previous sections can be dark matter, according to Fig.~\ref{fig:Omegachi}.
 
 By imposing the condition $\rho_{\rm fluid}/s<0.44$ eV, we find the upper bound on the reheating temperature as (the white region in Fig.~\ref{fig:TRmg1})
 \be
 T_R< 4.4 \times 10^3 \textrm{ GeV} \left( \frac{m_g}{\rm GeV} \right)^{1/3}\,,
 \label{eq:boundonTR}
 \ee
 so that the ratio $r_h/r_b$, for $T<T_R$, is bounded above by the tiny amount
 \be
 \frac{r_h}{r_b} < 6.1\times 10^{-12}\left( \frac{\rm GeV}{m_g}\right)^{1/3}\,,
 \ee
 which justifies the use of the graviton propagator in the absence of BH in section~\ref{sec:model} for the considered range of $m_g$. 
 
Within the considered range of values of $m_g$ in Tab.~\ref{tab:summary}, the typical values of the reheating temperature in Eq.~(\ref{eq:boundonTR}) and Fig.~\ref{fig:TRmg1} are relatively low, falling below, or borderline, the characteristic temperatures required by both thermal leptogenesis models $(T \gtrsim 10^9 - 10^{10} \, \textrm{GeV})$~\cite{Davidson:2002qv,Buchmuller:2004nz}  and the sphaleron mechanism  $(130 \, \textrm{GeV} \lesssim T \lesssim 10^{12} \, \textrm{GeV})$~\cite{Kuzmin:1985mm,Khlebnikov:1988sr,DOnofrio:2014rug}. However, some variants of these processes like low-scale leptogenesis can operate at temperatures close to the electroweak scale $(T \sim 10^2 - 10^3 \, \textrm{GeV})$~\cite{Akhmedov:1998qx,Pilaftsis:2003gt}, thus becoming compatible with the present reheating process.
 
On the other hand, there is an upper limit on supernovae (SN) emissivity at temperatures $T_{\rm SN}\simeq 30$ MeV given by~\cite{Hannestad:2007ys}
\be
\Delta_{\rm SN}\lesssim 3\times 10^{33} \textrm{erg cm}^{-3}\textrm{s}^{-1}\simeq 10^{-29} \textrm{ GeV}^5  \,.
\ee
We will only consider particles with masses $\lesssim 30$ MeV as the number density of heavier fields in thermal equilibrium in the SN should be very suppressed. This means considering mainly ($e^\pm,\nu's,\gamma$), for which
the values of the constants in Eq.~(\ref{eq:Delta_h}) become $c_{\rm SN}\simeq 1.94$ and $C_{\rm SN}\simeq 0.37$, leading to
\be
\Delta_{\rm SN}\simeq 2.4\times 10^{-13}\frac{\rm GeV^8}{M_5^3}\quad \Rightarrow\quad M_5\gtrsim 2.9\times 10^5 \textrm{ GeV}  \,,
\ee
 which is widely satisfied by the range in table~\ref{tab:summary}.

\section{Brane inflation}
\label{sec:brane_inflation}

In the case of brane cosmology, where matter fields are confined to the three-brane in the warped 5D space,  the Hubble parameter in the brane was computed for AdS spaces in Refs.~\cite{Shiromizu:1999wj,Binetruy:1999hy}. For the case of the linear dilaton background~\cite{Fichet:2022xol,Fichet:2023xbu,Fichet:2026nct},  the Hubble parameter can be written as
\be
3H^2M_4^2=\rho_b\left(1+\frac{\rho_b}{2\lambda}\right)+\rho_{\rm fluid},\quad \lambda=6 M_5^6/M_4^2,\quad \rho_{\rm fluid}=\frac{\lambda}{2}(r_h/r_b)^3  \,,
\ee
where $\rho_b$ is the brane energy density, $r_h$ the location of the horizon of the 5D black hole, and $\rho_{\rm fluid}$ a term, already introduced in section~\ref{subsec:fluid}, describing a holographic fluid which can play the role of holographic dark matter~\cite{Fichet:2026nct}, and whose strength depends on the size of the ratio $r_h/r_b$. As this term scales as matter, its strength will be negligible in the early universe, during inflation, where we expect that $\rho_b \gg 2\lambda$, and we can approximate the energy density by the inflaton $\phi$ potential $\rho_b\simeq V(\phi)$ so that we can approximate the Hubble by
\be
H^2\simeq \frac{V}{3M_4^2}\left(1+\frac{V}{2\lambda}  \right)\simeq \frac{V^2}{36 M_5^6}\,.
\label{eq:H-BC}
\ee

The slow-roll parameters for brane cosmology have been worked out in Refs.~\cite{Maartens:1999hf,Jaman:2018ucm}. For $V\gg 2\lambda$ they can be written as
\begin{align}
\epsilon(\phi)&\simeq 12 M_5^6 \frac{(V')^2}{V^3} \,, \\
\eta(\phi)&\simeq 12 M_5^6 \frac{V^{\prime\prime}}{V^2} \,, 
\end{align}
which yield the spectral tilt $n_s=1-6\epsilon+2\eta$. 

The end of inflation $\phi=\phi_f$ is fixed by the condition $\epsilon(\phi_f)=1$. The beginning of inflation $\phi=\phi_i$ is related to the number of e-folds before the end of inflation $N$ by
\be
N\simeq -\frac{1}{12 M_5^6}\int_{\phi_i}^{\phi_f}\frac{V^2}{V'}d\phi\,.
\ee
Finally, the CMB normalization of density perturbations yields
\be
A_s^2\simeq \frac{1}{(360\, \pi M_5^{9})^2}\frac{V^6(\phi_i)}{[V'(\phi_i)]^2} \simeq 2.1\times 10^{-9}  \,,
\ee
while the ratio of tensor-to-scalar perturbations~\cite{Jaman:2018ucm} is given by $r\simeq 24 \epsilon(\phi_i)$.

\subsection{A class of inflationary models}

Motivated by recent precision cosmological data from ACT
DR6~\cite{AtacamaCosmologyTelescope:2025blo}, and DESI
DR2~\cite{DESI:2025zgx} (signaling an upward shift in the scalar
spectral index and so a preference for deviations from pure
exponential plateau behavior),~\footnote{The present
  observation on DESI DR2 results evaluates specifically the scalar
  spectral index $n_s$, an analysis which is separate from DESI's
  high-profile constraints on dynamical dark energy, for which the
  DESI Collaboration is better known.}  it has been pointed out
recently, see e.g.~Ref.~\cite{Galiautdinov:2026jxi}, that polynomial
deformations of $\alpha$-attractor
models~\cite{Kallosh:2013lkr,Kallosh:2013hoa,Kallosh:2013yoa} should
have a better behavior when facing cosmological observables. As such,
a particular deformation of the $\alpha$-attractor potential, which
behaves as $(1-\mu^2/\phi^2)$ at high values of the field was proposed
for standard cosmology in Ref.~\cite{Galiautdinov:2026jxi}, while
being described by the $\alpha$-attractor model, $V_\alpha$, for low
values of $\phi$, in particular during reheating.

In this section we will study, for the sake of illustration, the case of brane (nonstandard) cosmology, a simpler potential, with essentially the same properties, given by
\be
V(\phi) = V_0 \times \left(\frac{\left(\frac{\phi}{\mu}\right)^2+\alpha}{\left(\frac{\phi}{\mu}\right)^2+1}-\alpha \right) \,, \label{eq:potential}
\ee
where $\alpha<1$ is a real parameter. The potential (\ref{eq:potential}) behaves in the different limits as
\begin{align}
V(\phi)&\simeq \frac{1}{2}m^2 \phi^2,\quad  \textrm{for}\quad \phi\ll\mu  \,, \nonumber\\
  V(\phi)&\simeq V_i\left( 1-\frac{\mu^2}{\phi^2} \right),\quad\textrm{with}\quad V_i=(1-\alpha)V_0=\frac{1}{2}m^2 \mu^2, \quad \textrm{ for}\quad \phi\gg\mu  \,.
\end{align}
These kinds of potentials are generic in string-based brane-world scenarios~\cite{Garcia-Bellido:2001lbk,Kachru:2003sx}.

In the UV regime, where $\phi/\mu\gg 1$, we get
\be
\epsilon\simeq \frac{96\mu^2 M_5^6}{m^2 \phi^6},\qquad \eta\simeq -\frac{144 M_5^6}{m^2\phi^4},\qquad N\simeq \frac{m^2}{192 M_5^6}\left(\phi_i^4-\phi_f^4 \right)  \,,
\ee
where $\phi_i$ and $\phi_f$ denote the values of the inflaton at the beginning and the end of inflation, respectively, and $N$ is the number of e-folds spanned.
In the approximation that $1\ll \phi_f/\mu\ll\phi_i/\mu$, we get
\be
\frac{\phi_f^6}{\mu^6}\simeq \frac{96 M_5^6}{m^2\mu^4} \,, \qquad \frac{\phi_i^4}{\mu^4}\simeq \frac{192 M_5^6}{m^2\mu^4}\,N  \,,
\ee
and the slow-roll parameters at the beginning of inflation $\phi_i$
\begin{equation}
\begin{aligned}
\epsilon_i&\simeq \frac{1}{16\sqrt{3}}\, \frac{m\mu^2}{M_5^3}\frac{1}{N^{3/2}}\quad \Rightarrow\quad r\simeq 24 \epsilon_i\simeq \frac{\sqrt{3}}{2}\, \frac{m\mu^2}{M_5^3}
\frac{1}{N^{3/2}} \,, \\
\eta_i&\simeq -\frac{3}{4N}\quad \Rightarrow\quad n_s\simeq 1-\frac{3}{2N} \,,
\end{aligned}
\end{equation}
where $r$ is the tensor-to-scalar ratio. Moreover, the value of $A_s^2$ is given~by
\be
A_s^2\simeq \frac{1}{1800\sqrt{3}\pi^2}\, \frac{m^5\mu^4}{M_5^9}N^{3/2}\simeq 2.1\times 10^{-9}  \,.
\label{eq:As2}
\ee

We can now particularize the previous results to our model, for which the inflaton mass is given in terms of $M_5$ as $m\simeq e^{5/8}M_5$, from Eq.~(\ref{eq:M5mexpansion}). Then the condition (\ref{eq:As2}) gives
\be
\mu\simeq 8.8\times 10^{-3} M_5 \left( \frac{60}{N}\right)^{3/8}  \,,
\label{eq:muM5}
\ee
and the values of $\phi_{i,f}$, for $N\simeq 60$, are given by
\be
\phi_f\simeq 0.36 M_5,\quad \phi_i\simeq 7.58 M_5  \,,
\ee
which satisfy the assumed condition $1 \ll \phi_{f}/\mu \ll \phi_{i}/\mu$, and thus evading the Lyth's bound~\cite{Lyth:1996im}. Finally the prediction for the tensor-to-scalar ratio is given by
\be
r\simeq 2.7\times  10^{-7}  \,.
\ee
Using now the relation (\ref{eq:muM5}) we can easily check that the initial condition $V(\phi_i) \gg 2\lambda$ implies the upper bound on $M_5$ (and so on $m_g$), as~\footnote{Use $M_5^3=\eta M_4^2=(2/3) m_g M_4^2$.}
\be
M_5\lesssim 3.4\times 10^{-3}M_4\simeq 8.2\times 10^{15}\textrm{ GeV} \,,
\ee
which is widely satisfied by our window in table~\ref{tab:summary}. Furthermore, the Hubble parameter during inflation is given by
\be
  H_i = 2.3 \times 10^{-5} M_5 \lesssim 4.5\times 10^{6} \textrm{ GeV}\,,
\ee
where the last inequality is coming from table~\ref{tab:summary}, and the maximum value of the reheat temperature $T_R^{\rm max}$, assuming good reheating, such that all the inflaton energy $V_i$ is converted into radiation, also gets the upper bound
\be
T_R^{\rm max}\lesssim  0.044 \, M_5\leq  8.9 \times 10^{9} \textrm{ GeV} \,.
\ee
Notice that this maximum value of the reheating temperature is larger than the upper bound of Eq.~(\ref{eq:boundonTR}), which is then permitted.

\subsection{Comments on the number of e-folds}

We will close this section with some comments on the observed number of e-folds $N_k$ when the pivot scale $k$ crosses the horizon.  On the one hand, for the case of low scale inflation $V_i\ll M_4^4$ and low reheating temperature $\rho_R\ll V_i$,  then $N_k$ tends to be reduced in standard cosmology (SC). On the other hand, for brane cosmology (BC), when $V_i\gg 2\lambda$ the number of e-folds $N_k$ is increased. In general we can write~\cite{Liddle:2003as,ParticleDataGroup:2024cfk,Martin:2010kz}
\be
N_k^\omega\simeq 67-\log\frac{k}{a_0H_0}+\Delta N^\omega, \quad \Delta N^\omega=\Delta N^{\omega}_{\rm SC}+\Delta N_{\rm BC}  \,.
\ee 
Assuming that the energy density at the end of inflation is dominated by the potential $\rho_f\simeq V_i$, and using for $V_i\gg 2\lambda$ that, in brane cosmology, the Hubble parameter is linear in the energy density, Eq.~(\ref{eq:H-BC}), we obtain
\begin{equation}
\begin{aligned}
\Delta N_{\rm SC}^\omega&\simeq \frac{1}{4}\log \frac{V_i}{M_4^4}+\frac{1-3\omega}{12(1+\omega)}\log \frac{\rho_R}{V_i}-\frac{1}{12}\log g_*  \,,  \nonumber\\
\quad \Delta N_{\rm BC}&=\frac{1}{2}\log \frac{V_i}{2\lambda}\,,
\end{aligned}
\end{equation}
where $\rho_R$ is the energy density at  reheating, and $\omega\equiv(P_\phi+\rho_\gamma/3)/(\rho_\phi+\rho_\gamma)$ characterizes the effective equation of state during reheating~\cite{Martin:2010kz}. Using now $a_0 H_0\simeq 2.3\times 10^{-4} \textrm{ Mpc}^{-1}$, we obtain for the pivot scale $k=0.05  \textrm{ Mpc}^{-1}$,
$N_{0.05}^\omega\simeq 61.8+\Delta N^\omega$.

We will now consider two typical benchmark points (BP) for two values of $m_g$ and the reheating temperature $T_R$: 
\begin{description}
\item[\textit{i)}] $m_g=1$ MeV and $T_R=200$ GeV,
\item[\textit{ii)}] $m_g=1$ TeV and $T_R=44.2$ TeV.  
\end{description}
For BP~(i) the long-lived isolated massive graviton can make the DM while the holographic fluid (see Sec.~\ref{sec:fluid}) is subleading. For BP~(ii) the massive graviton decays, while the holographic fluid can be the main component of DM. The different scales for both BPs are given in table~\ref{tab:scales} 
\begin{table}[h]
\centering
  \resizebox{15cm}{!}{
\begin{tabular}{||c||c|c|c|c||c|c|c||}
\hline
BP &$m_g$ [GeV] &$m$ [GeV] &$\mu$ [GeV]& $T_R$ [GeV] & $V_i^{1/4}$ [GeV] & $\rho_R^{1/4}$ [GeV] & $(2\lambda)^{1/4}$ [GeV]\\
\hline\hline
 \textit{(i)}& $10^{-3}$& $2.9\times 10^{11}$  &$1.4\times 10^9$ & 200& $1.7\times 10^{10}$ & 490 &$7.5\times 10^7$\\
 \hline
\textit{(ii)} & $10^3$ &$2.9\times 10^{13}$ &$1.4 \times 10^{11}$ & $4.42\times 10^4$ &$1.7\times 10^{12}$  & $1.1\times 10^5$ & $7.5\times 10^{10}$\\
 \hline\hline
\end{tabular}
}
\caption{\it Different scales for the considered BPs.}
 \label{tab:scales}
\end{table}
\begin{table}[h]
\centering
\begin{tabular}{||c|c|c|c|c|c|c|c||}
\hline
BP&$ \Delta N_{\rm SC}^0$ & $ \Delta N_{\rm SC}^{1/3}$ & $ \Delta N_{\rm BC}$& $N_{0.05}^0$& $N_{0.05}^{1/3}$ & $n_s^0$ & $n_s^{1/3}$ \\
\hline\hline
 \textit{(i)} & -24.9& -19.2& 10.8 & 47.7&53.5 &0.9685&0.9719\\
 \hline
 \textit{(ii)}  &-20.1  & -14.6 & 6.3 &48.0& 53.5 & 0.9688&0.9720 \\
 \hline\hline
\end{tabular}
\caption{\it Predictions for the different BPs.}
 \label{tab:efolds}
\end{table}
The predictions for both BPs are shown in table~\ref{tab:efolds} where 
$
n_s^\omega\simeq 1-\frac{3}{2 N_{0.05}^\omega}\,.
$
We have considered in table~\ref{tab:efolds} the two extreme cases where $\omega=0$ (1/3), so that the system of inflaton and matter between the end of inflation and reheating behaves as matter (radiation). We have to compare the predictions in table~\ref{tab:efolds} with the experimental values from Planck~\cite{Planck:2018jri}, $n_s=0.9649\pm 0.0042$, from ACT DR6~\cite{AtacamaCosmologyTelescope:2025blo}, $n_s=0.974\pm 0.003$, and from SPT-3G~\cite{SPT-3G:2025bzu}, $n_s=0.9679\pm 0.0033$.

Of course, establishing a realistic inflationary model consistent with all the required properties of our dark matter model is beyond the scope of the present paper.

 \section{Conclusion and outlook}
\label{sec:conclusion}

In this paper we have considered a brane-world 5D theory with a linear dilaton background, which amounts to a metric $A(z)$ linear in conformal coordinates, and with a holographic dual described by a little string theory. This theory appears to be an interesting alternative to the widely studied brane-world theories based on 5D AdS spaces, with a CFT as holographic dual. In fact, in the linear dilaton background, the brane cosmology contains a presureless fluid which can play the role of dark matter, while in AdS backgrounds the fluid is a dark radiation. Moreover, in both theories the bulk graviton propagator has a continuum of poles, a 5D realization of unparticles theories. However, while in the AdS background the continuum is gapless, in the linear dilaton background the continuum has a mass gap $m_g$ which is determined by the VEV on the brane of a stabilizing bulk field. This feature, which is essential in these types of theories, has been widely exploited in this paper.

The graviton propagator can be split into a massless term (describing the physical massless graviton) and a gapped continuum, which describes a continuum of poles from $s=m_g^2$ to $\infty$. As the physical graviton masslessness is protected by the diffeomorphism invariance, we have only considered the self-energy renormalization of the gapped continuum from its interaction with the energy-momentum tensor of Standard Model fields, an interaction which is suppressed by $1/M_5^{3/2}$.

One-loop radiative corrections to the gapped continuum propagator trigger an isolated resonance $\chi$ which manifests itself as a pole in the second Riemann sheet, corresponding to a resonance with a width: a good candidate to long-lived massive graviton. This phenomenon was already known in the past for the general case of gapped unparticle propagators, where it was shown that the mass of the isolated resonance is very close to the mass gap~\cite{Delgado:2008gj}. In our case, the mass of the massive graviton $m_\chi$ depends on the mass of particles contributing to the renormalization. Interestingly enough, we have found that the value of $m_\chi$ is dominated by the heaviest state contributing to the renormalization and, in particular, the presence of a scalar with a mass $m\simeq M_5$ triggers the value  $m_\chi\simeq m_g$. 

A comment about the calculation is in order here. When considering the contribution from gauge bosons, the Feynman gauge $\xi=1$ was chosen, while all integrals were computed in dimensional regularization, and the $\overline{\rm MS}$ renormalization scheme was chosen. While the gauge and renormalization scheme choices can affect the actual value of the self-energies, physical quantities extracted from the full renormalized theory, as the location of the pole mass and the pole width, are expected to be independent of the gauge-fixing parameter and the renormalization scheme, up to the order at which the perturbative calculation is performed.

A natural candidate for a scalar with a mass $m\simeq M_5$ is an inflaton with a potential localized on the brane. We have worked out a simple inflationary model that triggers inflation, as well as correct values of cosmological observables, in nonstandard cosmology where the evolution is triggered by a Hubble parameter $H\propto \rho_b$ (instead of $H\propto \rho_b^{1/2}$ as in standard cosmology). The CMB normalization, implying low scale inflation, along with the requirement of low reheating temperature tend to decrease the required number of e-folds. However there is an extra term in the number of e-folds arising from the brane cosmology which partially compensates the previous effect so that, for models with nonexponential flattening, the final prediction for the spectral index agrees with the recent measurements from ACT DR6 and DESI DR2.

The long-lived massive graviton is a candidate to feebly interacting dark matter, provided that a number of particle physics, cosmological and astrophysical constraints are fulfilled. 
As the massive graviton is feebly coupled with the Standard Model sector, with a coupling as $\mathcal L_{\rm int}=-(\lambda_\chi/M_4)\chi_{\mu\nu}T^{\mu\nu}$, where $\lambda_\chi$ is the Wilson coefficient of the effective operator, the former is not in thermal equilibrium with the SM plasma, and its energy density vanishes at the reheating temperature, after cosmological inflation, being generated by the freeze-in mechanism. We have adapted the mechanism studied in Ref.~\cite{Cai:2021nmk}, with channels producing $\chi$ involving the QCD coupling, as $q\bar q\to g\chi$ and $qg\to q\chi$, for heavy quarks $q=c,b$ and $g$ being the gluon, for which the freeze-in is IR dominated.
The final score is that the massive graviton with a mass in the range $m_\chi\in [0.02,2]$ MeV can satisfy all the constraints and have the right dark matter abundance depending on the value of the Wilson coefficient $\lambda_\chi$. 

\begin{figure}[htb]
\begin{center}  
  \includegraphics[width=0.8\textwidth]{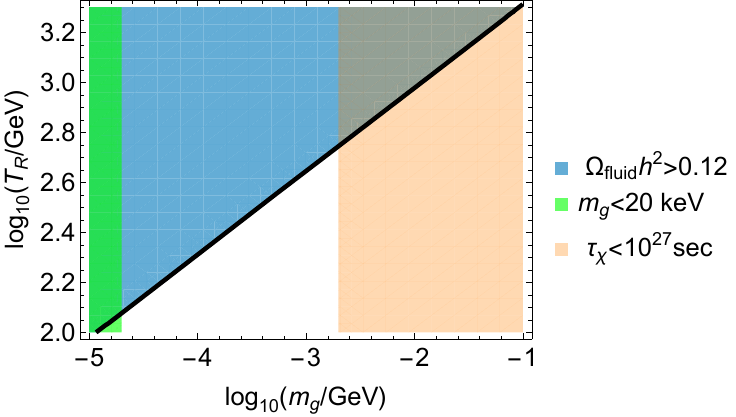}
\end{center}
     \caption{\it Summary plot in the plane $(m_g,T_R)$. In the white region the long-lived massive graviton is a candidate to dark matter, with an abundance which depends on the Wilson coefficient $\lambda_\chi$. In the orange shaded region the massive graviton is not allowed for dark matter as its lifetime is shorter than $\sim10^{27}$s. The blue shaded region is forbidden as there the holographic fluid would overclose the universe. Along the thick solid line the holographic fluid satisfies the required value for dark matter abundance.}
     \label{fig:TRmg}
 \end{figure}

On the other hand, the holographic fluid is always there, with an energy density overwhelming the critical value above the thick solid line of Fig.~\ref{fig:TRmg}, and subleading below it, which translates into a condition on the reheating temperature after inflation. For the considered mass range of the long-lived massive graviton $m_\chi\in[10\textrm{ keV},2\textrm{ MeV}]$, the condition for the fluid to not overclose the universe translates into the rough bound $T_R\lesssim 1$~TeV. For heavier masses $m_\chi\in[2 \textrm{ MeV},1\textrm{ GeV}]$, the massive graviton is stable but the region is forbidden by astrophysical data, while for $m_\chi\gtrsim 1$ GeV, the graviton is unstable on cosmological times and the region is permitted. In all cases  
the condition for the holographic fluid to have the dark matter abundance is given by the thick solid line in Fig.~\ref{fig:TRmg}, so that only the region below it satisfies the condition
$\Omega_{\rm fluid}h^2<0.12$ and there it is possible to assign dark matter to the massive long-lived graviton. Another interesting possibility in the white region is that dark matter has two components: a holographic fluid and long-lived massive gravitons with abundances such that $\Omega_{\rm fluid}h^2+\Omega_\chi h^2\simeq 0.12$.

In Ref.~\cite{Fichet:2026nct} phenomenological constraints were already imposed on the holographic fluid. It was found that the strongest one was the deviation from the Newton potential, imposing the lower bound $M_5\gtrsim 3\times 10^5$ TeV, widely satisfied in our model. On the other hand as long-lived massive gravitons are FIMPs, they should not be detected by direct detection experiments. Neither indirect experiments are expected to be able to detect massive gravitons. For instance the Bullet Cluster bounds the cross section for $\chi\chi\to\chi\chi$ to values $\sigma/m_\chi\lesssim 4\times 10^3 \textrm{ GeV}^{-3} $~\cite{Robertson:2016xjh}. However the FIMP is essentially collisionless as parametrically $\sigma/m_\chi\sim m_\chi/M_4^4$, many orders of magnitude below the Bullet Cluster limit. 

Confronting our long-lived FIMP with experimental data is one of the next challenges to be considered in he future.

\begin{acknowledgments}
  E.M. would like to thank the Instituto de F\'{\i}sica of the Universidade de S\~ao Paulo, Brazil, and the ICTP South American Institute for Fundamental Research (SAIFR), S\~ao Paulo, Brazil, for hospitality and financial support during the initial stages of this work. The works of E.M. and M.Q. are supported by the ``Proyectos de Investigaci\'on Precompetitivos'' Program of the Plan Propio de Investigaci\'on of the University of Granada under Grant No. PP2025PP-18. The research of M.Q. is also supported by the Grant No. PID2023-146686NB-C31 funded by MICIU/AEI/10.13039/501100011033/ and by ERDF/EU, and under Severo Ochoa Centres of Excellence Programme 2025-2029 (CEX2024001442-S). IFAE is partially funded by the CERCA program of the Generalitat de Catalunya.
\end{acknowledgments}

 \appendix
 
 \section{Calculation of the self-energy corrections}
 \label{sec:appendix}
 
In this Appendix we will provide some details on the one-loop corrections to the self-energy of the brane-to-brane propagator in the LD background, from the different fields $\varphi\varphi$ coupled to $h_{\mu\nu}(x)$. We will generically use the linear coupling as $\sim\kappa h \varphi\varphi$ which provides a correction to the self-energy by sunset diagrams which depend on the external momentum. Quadratic couplings $\sim\kappa^2 h h \varphi\varphi$ will contribute with seagull diagrams, which do not depend on the external momentum and thus only contribute to the mass renormalization, but do not contribute to the width. Therefore we will only compute this correction, in section~\ref{sec:appendix-scalar}, for the heaviest state, in particular a heavy scalar (the inflaton).

\subsection{Self-energy from scalars}
\label{sec:appendix-scalar}

For a real scalar $\phi$ with mass $m$, the energy-momentum tensor and the Feynman rule for the interaction of two scalars are given in Eqs.~(\ref{eq:tensor-scalar}) and (\ref{eq:coupling-scalar}), respectively.
The structure of the dressed propagator $\bar G_{\mu\nu;\alpha\beta}(p)$ is given by~Eq.~(\ref{eq:dressed}) with
\be
-i\Sigma^{\alpha'\beta';\mu'\nu'}=\frac{\kappa^2}{2}\int \frac{d^4q}{(2\pi)^4} i \Gamma^{\alpha'\beta'}(q,-q-p)i\Gamma^{\mu'\nu'}(-q,q+p) \frac{i}{q^2-m^2}\frac{i}{(q+p)^2-m^2}  \,,
\ee
where we have introduced a factor $1/2$ for identical fields and identified in the loop of scalars the momenta $p_1=q$ and $p_2=-q-p$, where $p$ is the external momentum. 

After introducing the Feynman parameter $x$, the previous integral can be transformed into
\be
-i\Sigma^{\alpha'\beta';\mu'\nu'}=\frac{\kappa^2}{2}\int\frac{d^4\ell}{(2\pi)^4}\int_0^1 dx \frac{ \Gamma^{\alpha'\beta'}(\ell-x p,-\ell-(1-x)p)\Gamma^{\mu'\nu'}(-\ell+x p,\ell+(1-x)p)}{\left(\ell^2-m^2+x(1-x)s\right)^2}  \,,
\ee
where we have used that $s=p^2$ and the change of variables $q=\ell-x p$. In the product
\be
i P_{\mu\nu;\alpha'\beta'}(-i\Sigma^{\alpha'\beta';\mu'\nu'})i P_{\mu'\nu';\alpha\beta}
\ee
the term in the vertices as $\eta^{\alpha'\beta'}$ and $\eta^{\mu'\nu'}$ vanish as the external propagators are traceless, as well as the terms proportional to the external momentum $p$ vanish by transversality of the external propagators. At the end we have 
\be
-i\Sigma^{\alpha'\beta';\mu'\nu'}=\frac{\kappa^2}{2}\int\frac{d^4\ell}{(2\pi)^4}\int_0^1 dx \frac{ 4 \ell^{\alpha'}\ell^{\beta'} \ell^{\mu'} \ell^{\nu'}}{\left(\ell^2-m^2+x(1-x)s\right)^2}  \,.
\label{eq:self-scalar}
\ee
Under the integral we can use the identity
\be
\ell^{\alpha'}\ell^{\beta'} \ell^{\mu'} \ell^{\nu'}=\frac{1}{d(d+2)}\left(\eta^{\alpha'\beta'}\eta^{\mu'\nu'}+\eta^{\alpha'\mu'}\eta^{\beta'\nu'}+\eta^{\alpha'\nu'}\eta^{\beta'\mu'}\right)(\ell^2)^2  \,,
\label{eq:4gammas}
\ee
where the first term does not contribute by the tracelessness property of the external propagators, while the second and third terms contribute equally by their symmetry properties. We then have that
\be
P_{\mu\nu;\alpha'\beta'}\left(\eta^{\alpha'\mu'}\eta^{\beta'\nu'}+\eta^{\alpha'\nu'}\eta^{\beta'\mu'}\right)P_{\mu'\nu';\alpha\beta}=2 P_{\mu\nu}^{\hspace{0.2cm}\mu'\nu'}P_{\mu'\nu';\alpha\beta}=2P_{\mu\nu:\alpha\beta}  \,,
\ee
and
\be
i P_{\mu\nu;\alpha'\beta'}(-i\Sigma^{\alpha'\beta';\mu'\nu'})i P_{\mu'\nu';\alpha\beta}=iP_{\mu\nu;\alpha\beta}\Sigma(s)  \,,
\ee
where 
\be
\Sigma = i\kappa^2 \int_0^1 dx\int \frac{d^d\ell}{(2\pi)^d}\frac{4}{d(d+2)}\frac{(\ell^2)^2}{(\ell^2-\Delta)^2},\quad \Delta=-x(1-x)s+m^2   
\ee
is the shift after resummation in the inverse propagator $G_h^{-1}\to G_h^{-1}-\Sigma$, i.e.
\be
\bar G_{\mu\nu;\alpha\beta}(p) = \frac{i P_{\mu\nu;\alpha\beta}}{G_h^{-1}-\Sigma}  \,.
\label{eq:dressed_final}
\ee

Using now dimensional regularization, we can write
\be
\Sigma = -\kappa^2\int_0^1 dx \frac{\Gamma(-d/2)}{(4\pi)^{d/2}}\left(\frac{1}{\Delta}\right)^{-d/2} \,. 
\ee
In the $\overline{\rm MS}$ renormalization scheme, the singularity 
\be
\frac{2}{4-d}-\gamma_E+\log 4\pi
\ee
is removed, and one gets the renormalized quantity
\be
\tilde\Sigma = -\frac{\kappa^2}{64\pi^2}\int_0^1 dx\,\Delta^2[3 - 2\log(\Delta/\mu^2) ]  \,,
\ee
where $\mu$ is the renormalization scale. The integral over $x$ can easily be done analytically, with the result 
\begin{align}
\widetilde\Sigma&=\frac{\kappa^2}{(120 \pi)^2}\left[30s^2\lambda^{5}(s) \log\left( \frac{1+\lambda(s)}{\sqrt{\lambda^2(s)-1}} \right) - 1155m^4 + 430m^2 s - 46s^2\right.\nonumber\\
&\left. + 15(30m^4-10m^2 s+s^2)\log(m^2/\mu^2)\right] \,,  \label{eq:SigmaR_scalar}
\end{align}
where $\lambda(s) = \sqrt{1-4 m^2/s}$. 

In the following we will consider the prescription $s \to s + i \epsilon$, and write the logarithmic term as
\be
30\lambda^5(s)\log\left( \frac{1+\lambda(s)}{\sqrt{\lambda^2(s)-1}} \right) = 15\lambda^5(s)\log\left( \frac{1+\lambda(s)}{1-\lambda(s)} \right) - 15\lambda^5(s)i\pi \,.
\ee
Therefore for the real ($\widetilde \Sigma_R$) and imaginary ($\widetilde\Sigma_I$) components of $\widetilde\Sigma$ we get
\begin{align}
  \widetilde\Sigma_R &= \frac{\kappa^2}{(120 \pi)^2}\left[ 30s^2\lambda^5 \mathcal F(\lambda) - 1155m^4 + 430m^2 s - 46s^2 \right. \nonumber \\
    &\left. + 15(30m^4-10m^2 s+s^2)\log(m^2/\mu^2)\right]  \,, \\
    \widetilde\Sigma_I &= - \frac{\kappa^2}{960 \pi } s ^2 \lambda^{5} \, \Theta(s - 4 m^2) \,, 
\end{align}
where 
\be
\mathcal F(\lambda) = \log \left( \frac{1+\lambda(s)}{\sqrt{\lambda^2(s)-1}} \right) \Theta(4m^2-s) + \log \left( \frac{1+\lambda(s)}{\sqrt{1-\lambda^2(s)}} \right) \Theta(s-4m^2)  \,.
\label{eq:functionF}
\ee

The self-energy in the second Riemann sheet $\widetilde\Sigma^{\rm II}$ is obtained from Eq.~(\ref{eq:SigmaR_scalar}) by making the replacement  $\lambda(s) \to -\lambda(s)$, and we obtain the result given in Eqs.~(\ref{eq:ReSigmaII-scalar}) and (\ref{eq:ImSigmaII-scalar}), where it has been assumed that $0 < -\Im s \ll \Re s$.

\subsubsection{The seagull contribution}
\label{sec:appendix-tadpole}

The quadratic interactions for gravitons come from the decomposition $g_{\mu\nu}=\eta_{\mu\nu}+\kappa h_{\mu\nu}$ and the use of expansions~\footnote{An on-shell massive graviton is traceless, and thus satisfies $h=0$.}~\cite{Choi:1994ax}
\begin{align}
g^{\mu\nu}&=\eta^{\mu\nu}-\kappa h^{\mu\nu}+\kappa^2 h^{\mu\alpha}h_{\alpha}^{\,\nu}+\cdots \,, \nonumber\\
\sqrt{-g}=&1+\frac{\kappa}{2}h+\frac{\kappa^2}{8} \mathcal H+\cdots,\quad h\equiv h^\mu_\mu\,,\quad \mathcal H\equiv h^2-2h_{\mu\nu}h^{\mu\nu} \,,
\end{align}
in the action
\be
S_\phi=\int d^4x \sqrt{-g}\left[\frac{1}{2}g^{\mu\nu}\partial_\mu\phi\partial_\nu\phi-\frac{1}{2}m^2\phi^2  \right] \,.
\ee
This gives rise to the interaction Lagrangian
\be
\mathcal L_{hh\phi\phi}= \kappa^2\left[ \frac{1}{16}\mathcal H (\partial_\mu\phi\partial^\mu\phi-m^2\phi^2)+\frac{1}{2}\partial_\mu\phi\partial_\nu\phi\left(h_\alpha^\mu h^{\nu\alpha}-\frac{1}{2}  h h^{\mu\nu}\right) \right]\,.
\ee
The Feynman rule in momentum space for the process $h_{\mu\nu}(k_1)h_{\rho\sigma}(k_2) \phi(p_1)\phi(p_2)$, where all momenta are incoming, is given by
\be
iV_{\mu\nu,\rho\sigma}(p_1,p_2)=i\kappa^2 \mathcal T_{\mu\nu,\rho\sigma}(p_1,p_2)  \,,
\ee
with
\begin{align}
\mathcal T^{\mu\nu,\rho\sigma}&=\frac{1}{4}(\eta^{\mu\rho}\eta^{\nu\sigma}+\eta^{\mu\sigma}\eta^{\nu\rho}-\eta^{\mu\nu}\eta^{\rho\sigma})(p_1\cdot p_2-m^2)\nonumber\\
&-\frac{1}{8}\eta^{\mu\nu}(p_1^\rho p_2^{\sigma}+p_1^\sigma p_2^\rho)-\frac{1}{8}\eta^{\rho\sigma}(p_1^\mu p_2^{\nu}+p_1^\nu p_2^\mu)\nonumber\\
&+\frac{1}{4}\left[ \eta^{\mu\rho}(p_1^\nu p_2^\sigma+p_1^\sigma p_2^\nu)+\eta^{\mu\sigma}(p_1^\nu p_2^\rho+p_1^\rho p_2^\nu) \right]\nonumber\\
&+\frac{1}{4}\left[ \eta^{\nu\rho}(p_1^\mu p_2^\sigma+p_1^\sigma p_2^\mu)+\eta^{\nu\sigma}(p_1^\mu p_2^\rho+p_1^\rho p_2^\mu) \right]  \,,
\end{align}
which is symmetric under $(\mu\nu)\leftrightarrow (\rho\sigma)$ as it should.

\begin{figure}[htb]
 \begin{center}
  \includegraphics[width=0.45\textwidth]{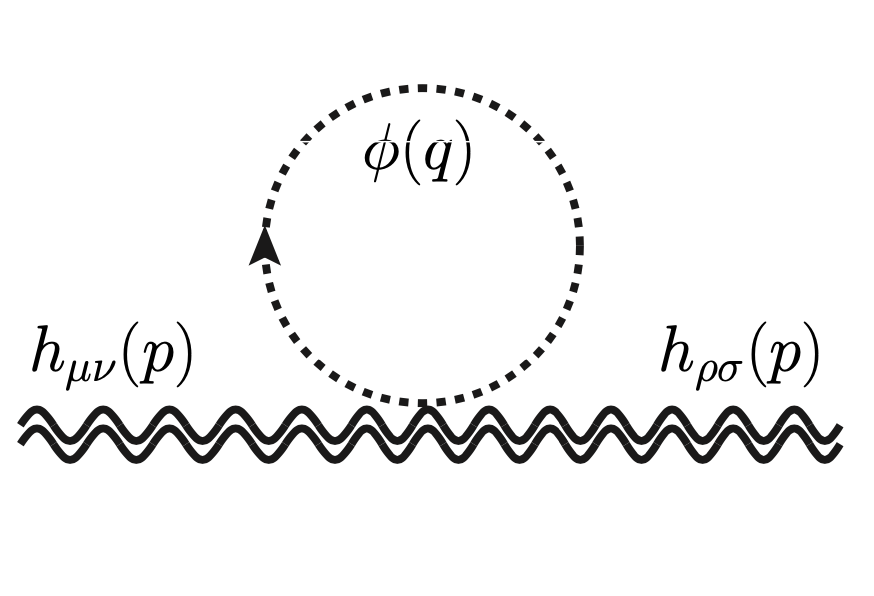} 
  \end{center}
\vspace{-0.7cm} \caption{\it The seagull diagram from scalars.}
     \label{fig:tadpole}
 \end{figure}

In the seagull diagram of Fig.~\ref{fig:tadpole}
the external graviton propagators, $P_{\alpha\beta,\mu\nu}$ and $P_{\rho\sigma,\gamma\delta}$ are traceless, so that the terms with $\eta^{\mu\nu}$ and $\eta^{\rho\sigma}$ vanish. Moreover, in the scalar loop the circulating momentum is $q$ (independent on the graviton momenta), so that $p_1=q$, $p_2=-q$, and
\begin{align}
\mathcal T_{\mu\nu,\rho\sigma}(q,-q)=&-\frac{1}{2}I_{\mu\nu,\rho\sigma}(q^2+m^2)\\
&-\frac{1}{2}(\eta_{\mu\rho}q_\nu q_\sigma+\eta_{\mu\sigma}q_\nu q_\rho+\eta_{\nu\rho}q_\mu q_\sigma+\eta_{\nu\sigma}q_\mu q_\rho) \,, \nonumber
\end{align}
where
\be
I_{\mu\nu,\rho\sigma}=\frac{1}{2}(\eta_{\mu\rho}\eta_{\nu\sigma}+\eta_{\mu\sigma}\eta_{\nu\rho})
\ee
satisfies the property $P_{\alpha\beta,\mu\nu}I^{\mu\nu,\rho\sigma}P_{\rho\sigma,\gamma\delta}=P_{\alpha\beta,\gamma\delta}$.

Under integration in dimensional regularization we can use the property $q_\mu q_\nu=(1/d)q^2 \eta_{\mu\nu}$ so that under integration we can write
\be
\mathcal T_{\mu\nu,\rho\sigma}=-I_{\mu\nu,\rho\sigma}\left[\frac{1}{2}(q^2+m^2)+\frac{2}{d}q^2 \right]  \,,
\ee
and, using dimensional regularization,
\be
-i\Sigma_{\mu\nu,\rho\sigma}=\kappa^2 \int \frac{d^dq}{(2\pi)^{d}}i\mathcal T_{\mu\nu,\rho\sigma}(q,-q)\frac{i}{q^2-m^2}  \,,
\ee
i.e.
\be
-i \Sigma=\kappa^2\int \frac{d^dq}{(2\pi)^{d}}\left[ \left( \frac{1}{2}+\frac{2}{d} \right)q^2+\frac{1}{2}m^2 \right]\frac{1}{q^2-m^2} \,.
\ee
Finally, using dimensional regularization integrals we can write
\be
\Sigma=\kappa^2 m^2\left(1+\frac{2}{d} \right)\frac{1}{(4\pi)^{d/2}}\Gamma\left( 1-\frac{d}{2}\right)\left(\frac{1}{m^2} \right)^{1-d/2} \,,
\ee
and, in the $\overline{\rm MS}$ renormalization scheme,
\be
\widetilde\Sigma=-\frac{1}{64\pi^2}\left(7+6\log\frac{\mu^2}{m^2} \right)\frac{m^4}{M_5^3} \,.
\ee

\subsection{Self-energy from fermions}
\label{sec:appendix-fermion}

We will here consider a Dirac fermion with mass $m$, energy-momentum tensor (\ref{eq:tensor-fermion}) and vertex coupling in momentum space (\ref{eq:coupling-fermion}).
The structure of the dressed propagator is given by Eq.~(\ref{eq:dressed}) with
\be
-i \Sigma^{\alpha'\beta';\mu'\nu'}=-\textrm{Tr}\int\frac{d^4q}{(2\pi)^4}i\Gamma^{\alpha'\beta'}(q,q+p)\frac{i(\slashed{p}+\slashed{q}+m)}{(q+p)^2-m^2} i\Gamma^{\mu'\nu'}(q+p,q)\frac{i(\slashed{q}+m)}{q^2-m^2}  \,,
\ee
where the global minus sign is from the fermion loop. Notice that the last term, as $\eta^{\mu\nu}$, in (\ref{eq:coupling-fermion}) does not contribute as the graviton propagator is traceless, while terms as $p^{\rho}$ where $(\rho=\alpha',\beta',\mu',\nu')$ do not contribute as the graviton propagator is transverse, so that effectively
\be
i\Gamma^{\mu\nu}(q,q+p)=i\Gamma^{\mu\nu}(q+p,q)=-i\frac{\kappa}{2}\left[\gamma^{\mu}q^\nu+\gamma^\nu q^\mu\right]=-i\kappa \gamma^\mu q^\nu  \,,
\label{eq:vertex-fermion2}
\ee
where in the last equality we have used the symmetry of the external propagator under
exchange of $\mu\nu$ indices.

We now introduce the Feynman parameter $x$ and using the change of variables $q=\ell -xp$ we get
\begin{align}
-i\Sigma^{\alpha'\beta';\mu'\nu'}&=-\kappa^2\int_0^1dx\int\frac{d^4\ell}{(2\pi)^4}\frac{\ell^{\alpha'}\ell^{\nu'}}{(\ell^2-\Delta)^2}A^{\beta'\mu'} \,, \nonumber\\
A^{\beta'\mu'}&=\textrm{Tr}\left[\gamma^{\beta'}[\slashed{A}+m] \gamma^{\mu'}[\slashed{B}+m]\right]  \,,
\end{align}
where we are defining
\be
A=\ell+(1-x)p,\quad B=\ell-xp  \,.
\ee

Using now the properties of the $\gamma$'s we can write
\be
A^{\beta'\mu'}=4(A^{\beta'}B^{\mu'}+A^{\mu'}B^{\beta'}-\eta^{\beta'\mu'}(A\cdot B-m^2)) \,,\quad
\ee
with
\be
A\cdot B=\ell^2+(1-2x)\ell\cdot p-x(1-x)p^2\equiv \ell^2+\Delta-m^2  \,, 
\ee
where we have removed the second term, as $\ell\cdot p$ is zero under integration in $\int d^4\ell$.
Finally using the transverse properties of external propagators we can write
\be
A^{\beta'\mu'}=4\left[2\ell^{\beta'}\ell^{\mu'}-\eta^{\beta'\mu'}(\ell^2+\Delta-2m^2) \right]   \,,
\ee
and
\be
-i\Sigma^{\alpha'\beta';\mu'\nu'}=-4\kappa_f^2 \int_0^1dx\int\frac{d^4\ell}{(2\pi)^4}
\frac{2\ell^{\alpha'}\ell^{\beta'}\ell^{\mu'}\ell^{\nu'}-\eta^{\beta'\mu'}\ell^{\alpha'}\ell^{\nu'}(\ell^2+\Delta-2m^2)  }{(\ell^2-\Delta)^2}   \,.
\ee

Using now the properties (\ref{eq:4gammas}) and
\be
\ell^{\alpha'}\ell^{\nu'}\equiv \frac{1}{d}\eta^{\alpha'\nu'}\ell^2
\label{eq:2gammas}
\ee
under dimensional regularization, and the tracelessness as well as the symmetric properties of the external propagators,
we get
\be
-i\Sigma^{\alpha'\beta';\mu'\nu'}=-4\kappa_f^2 \int_0^1dx\int\frac{d^d\ell}{(2\pi)^d}\frac{1}{(\ell^2-\Delta)^2}
\left[\frac{2-d}{d(2+d)}(\ell^2)^2-\frac{\Delta-2m^2}{d}\ell^2 \right] \eta^{\alpha'\mu'}\eta^{\beta'\nu'}  \,,
\ee
and then
\be
\Sigma=-4i\kappa_f^2 \int_0^1dx\int\frac{d^d\ell}{(2\pi)^d}
\left[\frac{2-d}{d(2+d)}\frac{(\ell^2)^2}{(\ell^2-\Delta)^2}-\frac{\Delta-2m^2}{d}\frac{\ell^2}{(\ell^2-\Delta)^2} \right]  \,.
\ee

Using now dimensional regularization we get
\be
\Sigma=\kappa^2\int_0^1 dx \frac{2}{(4\pi)^{d/2}}\Gamma(-d/2)\left[ 1-d+d\frac{m^2}{\Delta}\right]\left(\frac{1}{\Delta}\right)^{-d/2}  \,,
\ee
and in the $\overline{\rm MS}$ renormalization scheme
\be
\widetilde\Sigma=\frac{\kappa^2}{32\pi^2}\int_0^1 dx\,\Delta\left[8m^2-5\Delta+(6\Delta-8m^2)\log(\Delta/\mu^2) \right]   \,.
\ee
After integration over $x$ the result is
\begin{align}
\widetilde\Sigma&= - \frac{\kappa^2}{7200\pi^2}\left[-30 (8 m^2 + 3s) s\lambda^{3} \log\left( \frac{1+\lambda(s)}{\sqrt{\lambda^2(s)-1}}\right) - 1635 m^4 - 190 m^2 s+ 108 s^2\right.\nonumber\\
&\left.+15(30m^4+10m^2 s - 3s^2)\log(m^2/\mu^2) \right]  \,,
\end{align}
which yields
\begin{align}
\widetilde \Sigma_R&= - \frac{\kappa_f^2}{7200\pi^2}\left[-30 (8 m^2 + 3s) s\lambda^{3}\mathcal F(\lambda) - 1635 m^4 - 190 m^2 s+ 108 s^2\right.\nonumber\\
&\left.+15(30m^4+10m^2 s - 3s^2)\log\frac{m^2}{\mu^2}\right] \,, \\
\widetilde\Sigma_I&= -\frac{\kappa_f^2}{480\pi} (8 m^2 + 3s) s\lambda^{3}\,\Theta(s-4m^2)  \,,
\end{align}
and in the second Riemann sheet we get Eqs.~(\ref{eq:ReSigmaII-fermion}) and (\ref{eq:ImSigmaII-fermion}).

\subsection{Self-energy from the massive gauge sector}

 Here we consider the case of massive (i.e.~$W^\pm,Z$)  gauge bosons and the corresponding Faddeev-Popov ghosts.

 \subsubsection{Gauge bosons}
 
 \label{sec:appendix-massivegb}
 
We start with the Proca Lagrangian for a massive gauge boson with mass $m$, energy-momentum tensor given by Eq.~(\ref{eq:tensor-massivegb}) and vertex interaction in momentum space given by Eq.~(\ref{eq:coupling-massivegb}). Working in the Feynman gauge, the self-energy is given by
\begin{align}
-i\Sigma^{\alpha'\beta';\mu'\nu'}&=\frac{1}{2}\int\frac{d^4q}{(2\pi)^4}i\Gamma^{\alpha'\beta'}_{\quad\rho\sigma}(q,-q-p)\frac{-i\eta^{\rho\gamma}}{(q+p)^2-m^2}i\Gamma^{\mu'\nu'}_{\quad\gamma\delta}(-q,q+p)\frac{-i \eta^{\delta\sigma}}{q^2-m^2}\nonumber\\
&=\frac{1}{2}\int\frac{d^4q}{(2\pi)^4}\frac{\Gamma^{\alpha'\beta',\gamma\delta}(q,-p-q)\Gamma^{\mu'\nu'}_{\quad\gamma\delta}(-q,q+p)}{[(q+p)^2-m^2][q^2-m^2]}  \,,
\end{align}
where $1/2$ is a symmetry factor and, after introducing the Feynman parameter $x$ and making the change of variables $q=\ell-xp$ one gets
\be
-i\Sigma^{\alpha'\beta';\mu'\nu'}=\frac{1}{2}\int_0^1 dx \int\frac{d^d\ell}{(2\pi)^d}\frac{\Gamma^{\alpha'\beta',\gamma\delta}(q,-q-p)\Gamma^{\mu'\nu'}_{\quad\gamma\delta}(-q,q+p)}{(\ell^2-\Delta)^2}  \,.
\label{eq:self-GB-indices}
\ee 

Using the properties of external graviton propagators we obtain
\begin{align}
\Gamma^{\alpha'\beta',\gamma\delta}(p_1,p_2)/\kappa&=-\eta^{\gamma\delta}p_1^{\alpha'}p_2^{\beta'}+
(\eta^{\alpha'\gamma}p_1^\delta+\eta^{\alpha'\delta}p_1^\gamma)p_2^{\beta'}-\eta^{\alpha'\gamma}\eta^{\beta'\delta}K  \,, \\
\Gamma^{\mu'\nu'}_{\quad\gamma\delta}(-p_1,-p_2)/\kappa&=-\eta_{\gamma\delta}p_1^{\mu'}p_2^{\nu'}+
(\eta^{\mu'}_{\ \gamma}p_{1\delta}+\eta^{\mu'}_{\ \delta}p_{1\gamma})p_2^{\nu'}-\eta^{\mu'}_{\ \gamma}\eta^{\nu'}_{\ \delta}K  \,,
\end{align}
where 
\be
K=p_1\cdot p_2+2m^2,\quad p_1=q,\quad p_2=-q-p  \,.
\ee
The numerator of the self-energy (\ref{eq:self-GB-indices})
\be
N^{\alpha'\beta',\mu'\nu'}\equiv \Gamma^{\alpha'\beta',\gamma\delta}(p_1,p_2)\Gamma^{\mu'\nu'}_{\quad \gamma\delta}(-p_1,-p_2)
\ee
turns out to be
\begin{align}
N^{\alpha'\beta',\mu'\nu'}/\kappa^2&=2\left[\eta^{\alpha'\mu'}p_1^2\, p_2^{\beta'}p_2^{\nu'} +p_1^{\alpha'}p_2^{\beta'}p_1^{\mu'}p_2^{\nu'}-\eta^{\alpha'\mu'}(p_1^{\nu'}p_2^{\beta'}+p_1^{\beta'}p_2^{\nu'})K\right]+\eta^{\alpha'\mu'}\eta^{\beta'\nu'}K^2\nonumber\\
&=2\left[ \eta^{\alpha'\mu'}(q^2+2K)q^{\beta'}q^{\nu'}+q^{\alpha'}q^{\beta'}q^{\mu'}q^{\nu'}  \right]
+\eta^{\alpha'\mu'}\eta^{\beta'\nu'}K^2  \,,
\end{align}
where we have made use of the traceless and transverse properties of external propagators.

After the change of variables to $\ell=q+xp$ we obtain
\begin{align}
  N^{\alpha'\beta',\mu'\nu'}/\kappa^2&=2\left[\eta^{\alpha'\mu'} \ell^{\beta'}\ell^{\nu'}( \mathcal A - 2 (1-x) \ell^\alpha p_\alpha  -\ell^2) +\ell^{\alpha'}\ell^{\beta'}\ell^{\mu'}\ell^{\nu'}\right] \nonumber \\
  &+\eta^{\alpha'\mu'}\eta^{\beta'\nu'} \left[  (\mathcal B-\ell^2)^2 + (1-2x)^2 \ell^\alpha \ell^\beta p_\alpha p_\beta  - 2(1 -2x) (\mathcal B - \ell^2) \ell^\alpha p_\alpha  \right]\nonumber\\
\mathcal A&= x(2-x)p^2  +4m^2 \,,\nonumber\\ 
\mathcal B&=x(1-x) p^2  +2m^2  \,.
\end{align}
 
 Using all properties of symmetry of external propagators and those of integration under $d^d\ell$, the numerator of the self-energy (\ref{eq:self-GB-indices}) can be written as
 \be
 N^{\alpha'\beta',\mu'\nu'}=\eta^{\alpha'\mu'}\eta^{\beta'\nu'}N   \,,
 \label{eq:numerator-GB}
 \ee
 where 
 \begin{align}
 N/\kappa^2&=\frac{2\ell^2}{d}\left(\mathcal A-\ell^2\right)+\frac{4}{d(d+2)}(\ell^2)^2+(\mathcal B-\ell^2)^2+\frac{(1-2x)^2}{d}\ell^2 p^2 \nonumber\\
 &=\mathcal B^2+\left[\frac{2\mathcal A}{d}-2\mathcal B+\frac{(1-2x)^2}{d}p^2\right]\ell^2+\frac{d}{d+2}(\ell^2)^2  \,.
 \end{align}
 
 An obvious conclusion from the structure of Eq.~(\ref{eq:numerator-GB}) is that one recovers the general structure for the massive graviton propagator and the scalar shift is
 \be
\Sigma=\frac{i}{2}\int_0^1 dx \int\frac{d^d\ell}{(2\pi)^d}\frac{N}{(\ell^2-\Delta)^2} \,. \label{eq:self-GB-indices2}
\ee 
After integration over $\int d^d\ell$ one gets, using the dimensional regularization formalism,
\be
\Sigma=-\frac{1}{2}\int_0^1 dx\,\frac{\Gamma(2-d/2)}{(4\pi)^{d/2}}\left(\frac{1}{\Delta}\right)^{2-d/2}\mathcal N  \,,
\ee
where
\be
\mathcal N= \frac{\kappa^2}{d-2}\left[d (m^2+2x(1-x)s)^2 + (1-16x+18x^2)m^2 s - x(1-x)(1+2x)s^2
\right]  \,,
\ee
and in the $\overline{\rm MS}$ renormalization scheme,
 \begin{align}
\widetilde \Sigma&=\frac{\kappa^2}{64\pi^2}\int_0^1 dx\left\{2(m^2 + 2x(1-x)s)^2\right.\\
 & - \left.\left[4m^4 + (1 + 2x^2) m^2 s - x(1-x)(1 - 14x + 16x^2) s^2 \right](1-\log(\Delta/\mu^2)) \right\} \,. \nonumber
  \end{align}
After making the integral in $x$, the result turns out to be
  \begin{align}
 \widetilde   \Sigma &= \frac{\kappa^2}{(120 \pi)^2} \Bigg[ 15 (60 m^4 + 25 m^2 s + 3 s^2) \log(m^2/\mu^2) - 2 (1695 m^4 + 455 m^2 s + 24 s^2) \nonumber \\
      & + 30 (98 m^4 + 31 m^2 s + 3 s^2) \lambda(s) \log\left(\frac{1+\lambda(s)}{\sqrt{\lambda^2(s)-1}}\right) \Bigg] \,,  \label{eq:SigmaR_gb}
  \end{align}
so that
\begin{align}
\widetilde \Sigma_R&= \frac{\kappa^2}{(120 \pi)^2} \Bigg[ 15 (60 m^4 + 25 m^2 s + 3 s^2) \log(m^2/\mu^2) - 2 (1695 m^4 + 455 m^2 s + 24 s^2) \,, \nonumber \\
      & + 30 (98 m^4 + 31 m^2 s + 3 s^2) \lambda(s) \mathcal F(\lambda )\Bigg] \,, \\
\widetilde \Sigma_I&= -\frac{\kappa^2}{960\pi}(98 m^4 + 31 m^2 s + 3 s^2) \lambda(s)\, \Theta(s-4m^2)  \,,
\end{align}
and in the second Riemann sheet we recover expressions (\ref{eq:ReSigmaII-massivegb}) and (\ref{eq:ImSigmaII-massivegb}).

\subsubsection{Faddeev-Popov ghosts}

\label{sec:appendix-massiveFP}

The energy-momentum tensor and vertex to the massive graviton for FP fields $c$ and $\bar c$ with incoming momenta $p_1$ and $p_2$ are, in the Feynman gauge for which its mass is equal to the gauge boson mass $m_c=m$, given by Eqs.~(\ref{eq:tensor-massiveFP}) and (\ref{eq:coupling-massiveFP}).
Its contribution to the self-energy is given by
\be
-i\Sigma^{\alpha'\beta';\mu'\nu'}=- \int\frac{d^4q}{(2\pi)^4}i\Gamma^{\alpha'\beta'}(q,-q-p)\frac{i}{(q+p)^2-m^2}i\Gamma^{\mu'\nu'}(-q,q+p)\frac{i}{q^2-m^2}  \,,
\ee
where the global minus sign comes from the fact that the ghost scalars are anticommuting.

Using now the tracelessness and  symmetry properties of the external graviton propagators we can write
\be
i\Gamma^{\mu\nu}(p_1,p_2)\equiv 2 i \kappa\, p_1^\mu p_2^\nu   \,,
\ee
and moreover using the transverse properties of graviton propagators
\be
i\Gamma^{\mu\nu}(q,-q-p)=i\Gamma^{\mu\nu}(-q,q+p)=-2i\kappa q^\mu q^\nu   \,.
\ee

After introducing the Feynman parameter $x$ and using once again the property of transverse graviton propagators, we get
\be
-i\Sigma^{\alpha'\beta';\mu'\nu'}=-4\kappa^2\int_0^1 dx\int\frac{d^4\ell}{(2\pi)^4} \frac{\ell^{\alpha'}\ell^{\beta'}\ell^{\mu'}\ell^{\nu'}}{(\ell^2-\Delta)^2}   \,.
\ee

As expected the result is $-2$ times that of a real scalar in Eq.~(\ref{eq:self-scalar}), where the sign comes from the fact that the FP fields anticommute and the factor of 2 from the fact that FP fields are complex. Therefore we can write for the final renormalized result
\be
\widetilde\Sigma = \frac{\kappa^2}{32\pi^2}\int_0^1 dx\,\Delta^2[3 - 2\log(\Delta/\mu^2)]  \,.
\ee

After integration over $x$ the result is
\begin{align}
\widetilde\Sigma&=\frac{\kappa^2}{7200\pi^2}\left[-30s^2\lambda^{5}\log\left(\frac{1+\lambda(s)}{\sqrt{\lambda^2(s)-1}} \right) +1155m^4-430m^2 s+46s^2\right.\nonumber\\
&\left.-15(30m^4-10m^2 s+s^2)\log(m^2/\mu^2)\right]  \,,
\end{align}
and so
\begin{align}
\widetilde \Sigma_R&=\frac{\kappa^2}{7200\pi^2}\left[-30s^2\lambda^{5}\mathcal F(\lambda)+1155m^4-430m^2 s+46s^2\right.\nonumber\\
&\left.-15(30m^4-10m^2 s+s^2)\log(m^2/\mu^2)\right]  \,, \\
\widetilde \Sigma_I&= \frac{\kappa^2}{480\pi}s^2\lambda^{5}\,\Theta(s-4m^2) \,.
\end{align}
Finally, in the second Riemann sheet we recover the result from Eqs.~(\ref{eq:ReSigmaII-massiveFP}) and (\ref{eq:ImSigmaII-massiveFP}).

\subsection{Self-energy from the massless gauge sector}
 Here we consider the case of massless (i.e.~$\gamma,\, g^A$) gauge bosons and the corresponding Faddeev-Popov ghosts.
 
 \subsubsection{Gauge bosons}
 
 \label{sec:appendix-masslessgb}

   The interaction of massive gravitons with massless gauge bosons is given by Eq.~(\ref{eq:coupling-massive-gauge}) with $m=0$. The final result can also be obtained by taking the limit $m \to 0$ in Eq.~(\ref{eq:SigmaR_gb}), and using 
   \begin{equation}
    \lim_{m \to 0} \log\left( \frac{1 + \lambda(s)}{\sqrt{\lambda^2(s) - 1}} \right) = \frac{1}{2} \log(-s/m^2) \,,
   \end{equation}
   when considering the principal branch of the square root.  The renormalized self-energy is then given by
 \be
\widetilde \Sigma=-\frac{\kappa^2}{4800\pi^2}s^2\left[16-15\log(-s/\mu^2)  \right]  \,,
 \ee
so that
 \begin{align}
  \widetilde\Sigma_R &=  -\frac{\kappa^2}{4800\pi^2}s^2\left[16-15\log(s/\mu^2)  \right]   \,, \label{eq:Re_gb_massless} \\
  \widetilde\Sigma_I &= -\frac{\kappa^2}{320 \pi } s^2 \,\Theta(s-4m^2)   \,.  \label{eq:Im_gb_massless}
 \end{align}
 The self-energy in the second Riemann sheet is then given by Eqs.~(\ref{eq:ReSigmaII-masslessgb}) and (\ref{eq:ImSigmaII-masslessgb}).

\subsubsection{Faddeev-Popov ghosts}

\label{sec:appendix-masslessFP}

For the FP ghosts corresponding to the massless gauge fields  the result is 
\be
\widetilde\Sigma=\frac{\kappa^2}{7200\pi^2} s^2[46-15\log(-s/\mu^2)] \,. 
\ee
so that
\begin{align}
\widetilde \Sigma_R  &=  \frac{\kappa^2}{7200\pi^2} s^2[46-15\log(s/\mu^2)] \,, \\
\widetilde \Sigma_I  &= \frac{\kappa^2}{480\pi} s^2    \,,
\end{align}
and, in the second Riemann sheet, we recover the expressions (\ref{eq:ReSigmaII-masslessFP}) and (\ref{eq:ImSigmaII-masslessFP}).

\bibliographystyle{JHEP}
\bibliography{biblio}

\end{document}